\documentclass[reprint, aip, jcp, letterpaper, superscriptaddress]{revtex4-2}

\usepackage[T1]{fontenc} 

\usepackage{amsmath,color}
\usepackage{amssymb}
\usepackage{amsfonts}
\usepackage{amsthm}
\usepackage{mathtools}

\usepackage{algorithm}
\usepackage{algpseudocode}
\usepackage{dsfont}

\usepackage{bbold}
\usepackage{chemformula}
\usepackage{siunitx}
\usepackage{adjustbox}

\DeclarePairedDelimiterX\braket[2]{\langle}{\rangle}{#1 \delimsize\vert #2}
\DeclarePairedDelimiterX\braket3[3]{\langle}{\rangle}{#1 \delimsize\vert #2 \delimsize\vert #3}

\usepackage{accents}
\newcommand{\dbtilde}[1]{\accentset{\approx}{#1}}

\newcommand{\vB}{\mathbf{B}}

\newcommand{\vD}{\mathbf{D}}
\newcommand{\ve}{\mathbf{e}}

\newcommand{\vnabla}{\boldsymbol{\nabla}}
\newcommand{\vdelta}{\boldsymbol{\delta}}
\newcommand{\vmu}{\boldsymbol{\mu}}

\newcommand{\vf}{\mathbf{f}}
\newcommand{\vR}{\mathbf{r}}

\newcommand{\vk}{\mathbf{k}}

\newcommand{\vxi}{\boldsymbol{\xi}}

\newcommand{\hH}{\hat{H}}

\newcommand{\hP}{\hat{\boldsymbol{\mathcal{P}}}}

\newcommand{\avg}[1]{\left\langle #1\right\rangle}

\begin{document}

    \title{Mesoscale Molecular Simulations of Fabry--P\'erot Vibrational Strong Coupling}

\author{Tao E. Li}

\email{taoeli@udel.edu}

\affiliation{Department of Physics and Astronomy, University of Delaware, Newark, Delaware 19716, USA}


	\begin{abstract}
        Developing theoretical frameworks for vibrational strong coupling (VSC) beyond the single-mode approximation is crucial for a comprehensive understanding of experiments with planar Fabry--P\'erot cavities. Herein, a generalized cavity molecular dynamics (CavMD) scheme is developed to simulate VSC of a large ensemble of realistic molecules coupled to an arbitrary 1D or 2D photonic environment. This approach is built upon the Power--Zienau--Woolley Hamiltonian in the normal mode basis and uses a grid representation of the molecular ensembles to reduce the computational cost. When simulating the polariton dispersion relation for a homogeneous distribution of molecules in planar Fabry--P\'erot cavities, our data highlight the importance of preserving the in-plane translational symmetry of the molecular distribution. In this homogeneous limit, CavMD yields the consistent polariton dispersion relation as an analytic theory, i.e., incorporating many cavity modes with varying in-plane wave vectors ($k_{\parallel}$) produces the same spectrum as the system with a single cavity mode. Furthermore, CavMD reveals that the validity of the single mode approximation is challenged when nonequilibrium polariton dynamics are considered, as  polariton-polariton scattering occurs between  modes with nearest neighbor $k_{\parallel}$. The procedure for numerically approaching the macroscopic limit is also demonstrated with CavMD by increasing the system size. Looking forward, our generalized CavMD approach may facilitate understanding vibrational polariton transport and condensation.
	\end{abstract}
	
	\maketitle

        \section{Introduction}

        Vibrational strong coupling (VSC) has recently attracted great attention due to its potential for modifying molecular properties with vacuum electromagnetic fields. \cite{Shalabney2015,Long2015,Thomas2016, Dunkelberger2016, Thomas2019_science,Xiang2020Science,Chen2022,Ahn2023Science,Wright2023} Most VSC experiments to date have been performed in planar Fabry--P\'erot cavities, in which the Rabi splitting between an infrared (IR) cavity mode and a vibrational mode of a large collection of molecules is observed in the linear IR spectrum.\cite{Ribeiro2018,Li2022Review,Fregoni2022,Simpkins2023,Mandal2023ChemRev,Ruggenthaler2023} Alongside the exciting yet sometimes controversial experimental observations, \cite{Imperatore2021,Wiesehan2021,Fidler2023}  theoretical work on VSC \cite{Galego2019,Campos-Gonzalez-Angulo2019,Hoffmann2020,LiHuo2021,Fischer2021,YangCao2021,Wang2022JPCL,Flick2017,Riso2022,Schafer2021,Bonini2021,Yang2021,Rosenzweig2022,Triana2020Shape,Li2020Water,Haugland2020,Philbin2022} usually builds upon different levels of approximations; the most pervasive approximation being that only a single or a few cavity modes are taken into account. However, as will be discussed in detail in this work, because planar Fabry--P\'erot cavities contain a continuum of cavity modes along the cavity mirror plane, \cite{Li2022Review} the single-mode approximation might break down. 
        
        For example, under electronic strong coupling, experimentally observed exciton-polariton condensates  originate from elastic scattering between lower polaritons at different in-plane wave vectors (as denoted by $k_{\parallel}$). \cite{Deng2010,Keeling2020} In other words, including cavity modes at different $k_{\parallel}$ values is necessary for a proper description of exciton-polariton condensates. As another example, Ebbesen and coworkers observed that VSC catalytic effects occur only when the cavity mode at $k_{\parallel} = 0$ is at resonance with a vibrational mode of molecules. \cite{Thomas2016, Thomas2019_science}  
        These examples make it clear that the single-mode approximation is insufficient for describing these strong coupling effects in Fabry–Pérot cavities, and  the full cavity dispersion relation must be considered.

        Some recent studies have employed numerical simulations to investigate electronic strong coupling in multi-mode one-dimensional (1D) Fabry--P\'erot cavities. \cite{Luk2017,Tichauer2021,Sokolovskii2022,Sukharev2023} Under VSC conditions, however, most theoretical approaches have relied on the single-mode approximation, with an explicit treatment of the multi-mode Fabry--P\'erot cavities discussed only for model systems. \cite{Suyabatmaz2023,Ying2023} It is thus necessary to extend theoretical approaches for VSC, including electronic structure theory and  molecular dynamics simulations, to the multi-mode case.  Moreover, only by developing an adequate theoretical description of the multi-mode cavities, one can examine whether conclusions made in the single-mode limit still hold for more realistic cavity setups.
        
        One efficient simulation tool for describing VSC in the single-mode limit is the  cavity molecular dynamics (CavMD) approach, \cite{Li2020Water} in which the dynamics of a large ensemble of molecules coupled to a single cavity mode (or a few cavity modes) are propagated classically or semiclassically (via path integrals) on the electronic ground state surface. Some notable results of CavMD when simulating collective VSC are worth mentioning. (i) Under thermal equilibrium, cavity modifications of individual molecular properties become negligible even when nuclear and photonic quantum effects are considered. \cite{Li2020Water,Li2022RPMDCav} (ii) In the linear response limit, the polariton spectrum described by CavMD is  equivalent to that described by coupled oscillator models. \cite{Li2020Water,Lieberherr2023CavMD} (iii) In the weak excitation limit, the relaxation rates of the upper polariton (UP)  have the same parameter dependence as the Fermi's golden rule rate of a tight-binding harmonic model. \cite{Li2021Relaxation} (iv) In strong pumping regime, CavMD predicts novel mechanisms, including lower-polariton (LP) enhanced molecular nonlinear absorption, \cite{Li2020Nonlinear} a behavior that has been observed experimentally, \cite{Xiang2021JCP}  as well as strong energy accumulation in  solute molecules by exciting the solvent LP. \cite{Li2021Solute} The CavMD approach has also recently embraced a quantum mechanical/molecular mechanical (QM/MM) description of the molecular systems. This advance enables the direct simulation of ultrafast VSC experiments such as the pseudorotation ``reaction'' of the \ch{Fe(CO)5} solute in the \textit{n}-dodecane solvent, \cite{Li2023QMMM} although more extensive simulations (i.e., by including a large number of replicas of the solute-solvent molecular systems) are needed to understand the mechanisms of the ultrafast experiments. 

        In this manuscript, we extend the CavMD approach to describe VSC in arbitrary cavity setups, especially Fabry--P\'erot cavities with $k_{\parallel}$-dependent cavity modes.  We perform numerical experiments to highlight the importance of the in-plane translational symmetry in reproducing the $k_{\parallel}$-dependent polariton branches observed in experiments. Additionally, utilizing nonequilibrium CavMD simulations, we provide initial data on polariton-polariton scattering with neighboring $k_{\parallel}$ values, a fundamental mechanism which is relevant to the formation of polariton condensates.  Finally, our numerical experiments suggest research opportunities of preparing VSC in the absence of the molecular in-plane translational symmetry.

        The manuscript is organized as follows. In Sec. \ref{sec:theory}, we  present a derivation of CavMD in arbitrary photonic environments, especially in Fabry--P\'erot-like cavities. Sec. \ref{sec:simulation_details} provides simulation details. In Sec. \ref{sec:results}, we present CavMD results  for a few cavity setups. In Sec. \ref{sec:discussion}, we discuss the importance of the in-plane translational symmetry and the procedure to approach the macroscopic limit using CavMD. We conclude in Sec. \ref{sec:conclusion}.

        \section{Theory}\label{sec:theory}

        \subsection{QED Hamiltonian in a photonic structure} 

        We start with the Power--Zienau--Woolley Hamiltonian for light-matter interactions: \cite{Cohen-Tannoudji1997}
        \begin{equation}
            \hH_{\rm QED} = \hH_{\rm M} + \hH_{\rm ph} + \hH_{\rm int} .
        \end{equation}
        Here, $\hH_{\rm M}$ denotes the standard molecular (kinetic + potential) Hamiltonian; $\hH_{\rm ph}$ denotes the photonic Hamiltonian:
        \begin{equation}\label{eq:H_photon_first_quant}
            \hH_{\rm ph} = \frac{1}{2}\int d\vR\left ( \frac{1}{\varepsilon_0} |\hat{\vD}_{\perp}(\vR)|^2 + \frac{1}{\mu_0} |\hat{\vB}(\vR)|^2\right ) ,
        \end{equation}
        where $\hat{\vD}_{\perp}(\vR)$ and $\hat{\vB}(\vR)$ are the  displacement and magnetic field operators, and $\varepsilon_0$ and $\mu_0$ denote the vacuum permittivity and permeability, respectively; $\hH_{\rm int}$ denotes the  Hamiltonian that governs the light-matter interaction:
        \begin{equation}\label{eq:H_int}
            \hH_{\rm int} = - \frac{1}{\varepsilon_0}\int d\vR \hat{\vD}_{\perp}(\vR)\cdot \hP_{\perp}(\vR) + \frac{1}{2\epsilon_0} \int d\vR |\hP_{\perp}(\vR)|^2 .
        \end{equation}
        Here, the transverse polarization density operator $\hP_{\perp}(\vR)$ is defined as \cite{Cohen-Tannoudji1997}
        \begin{equation}\label{eq:P_perp_operator}
            \hP_{\perp}(\vR) = \int d\vR' \overleftrightarrow{\vdelta}_{\perp}(\vR - \vR') \hP(\vR') ,
        \end{equation}
        where the polarization density operator $\hP(\vR)$ reads
        \begin{equation}\label{eq:P_operator}
            \hP(\vR) = \sum_{i} \mathcal{Q}_i \hat{\vR}_i \delta(\vR - \vR_i) .
        \end{equation}
        In Eq. \eqref{eq:P_operator}, $\mathcal{Q}_i$ denotes the charge of the $i$-th particle, which can be either an electron or a nucleus; $\hat{\vR}_i$ denotes the position operator of the particle. In Eq. \eqref{eq:P_perp_operator},  the transverse $\delta$-function $\overleftrightarrow{\vdelta}_{\perp}(\vR - \vR')$ is a rank-two tensor:
        \begin{equation}
            \overleftrightarrow{\vdelta}_{\perp}(\vR - \vR')  = \sum_{k\lambda} \vf_{k\lambda}(\vR)\vf^{\ast}_{k\lambda}(\vR') .
        \end{equation}
        Here, $\vf_{k\lambda}(\vR)$ denotes the mode function of the photon with wave vector $k \equiv |\vk|$ and two possible polarization directions indexed by $\lambda$, and $\vf^{\ast}_{k\lambda}(\vR)$ denotes its complex conjugate. The explicit form of $\vf_{k\lambda}(\vR)$ depends on the boundary conditions of the dielectric medium; it can be obtained by solving the Helmholtz equation $\nabla\times\nabla\times\vf_{k\lambda}-k^2\vf_{k\lambda}=0$. For example, in vacuum, $\vf_{k\lambda}(\vR) = \vxi_{\lambda} e^{i\vk\cdot \vR}/\sqrt{\Omega}$, where $\vxi_{\lambda}$ denotes the unit vector along the photon polarization direction, and $\Omega$ denotes the quantization volume.

        \subsubsection{Normal mode representation}
        The field operators in Eqs. \eqref{eq:H_photon_first_quant} and \eqref{eq:H_int} can also be represented in the normal mode representation.
        In the normal mode representation, the displacement field operator $\hat{\vD}_{\perp}(\vR)$ reads
        \begin{equation}
            \hat{\vD}_{\perp}(\vR) = \hat{\vD}(\vR) = \sum_{k\lambda} 
            iD_{k}
            \left [ \hat{a}_{k\lambda} \vf_{k\lambda}(\vR)  - \hat{a}^{\dagger}_{k\lambda} \vf_{k\lambda}^{\ast}(\vR) \right ] .
        \end{equation}
        Here, $D_{k} = \sqrt{\hbar\omega_{k}\epsilon_0/2}$,  where $\omega_{k}$ denotes the photonic frequency, and $\hat{a}^{\dagger}_{k\lambda}$ ($\hat{a}_{k\lambda}$) denotes the creation (annihilation) operator of the photon mode. In Eq. \eqref{eq:H_photon_first_quant}, the magnetic field operator is given by 
        $\hat{\vB}(\vR) = \frac{1}{\sqrt{\epsilon_0}}\vnabla\times \hat{\mathbf{A}}(\vR)$.
        Here, the vector potential $ \hat{\mathbf{A}}(\vR)$ is 
        \begin{equation}
        \begin{aligned}
            \hat{\mathbf{A}}_{\perp}(\vR) & = \sum_{k\lambda} 
            A_{k}
            \left [ \hat{a}_{k\lambda} \vf_{k\lambda}(\vR)  + \hat{a}^{\dagger}_{k\lambda} \vf_{k\lambda}^{\ast}(\vR) \right ],
            \end{aligned}
        \end{equation}
        where $A_{k} =  \sqrt{\hbar/2\omega_k\epsilon_0}$.
        Hence, in the normal mode representation, the photonic Hamiltonian in Eq. \eqref{eq:H_photon_first_quant} can be expressed as
        \begin{equation}\label{eq:H_ph_second_quant}
            \hH_{\rm ph} = \sum_{k\lambda} \hbar\omega_{k}\hat{a}_{k\lambda}^{\dagger}\hat{a}_{k\lambda} .
        \end{equation}
        
        Given a collection of $N$ charge-neutral molecules, let $\hat{\vmu}^{(n)}$ be the molecular dipole operator of the $n$-th molecule. The  polarization density operator $\hP(\vR)$ in Eq. \eqref{eq:P_operator} can be rewritten as
        \begin{equation}\label{eq:P_operator_dipoles}
            \hP(\vR) = \sum_{n=1}^{N} \hat{\vmu}^{(n)} \delta(\vR - \vR_n) ,
        \end{equation}
        where $\vR_n$ denotes the position of the $n$-th molecule. When Eq. \eqref{eq:P_operator_dipoles} is substituted into the light-matter interaction Hamiltonian defined in Eq. \eqref{eq:H_int}, $\hH_{\rm int}$ becomes \footnote{
        To derive Eq. \eqref{eq:H_int_normal_mode} from Eq. \eqref{eq:H_int}, we have used the following identities:  $- \frac{1}{\varepsilon_0}\int d\vR \hat{\vD}_{\perp}(\vR)\cdot \hP_{\perp}(\vR)  = - \frac{1}{\varepsilon_0}\int d\vR \hat{\vD}_{\perp}(\vR) \cdot \hP(\vR)$, and $\frac{1}{2\epsilon_0} \int d\vR |\hP_{\perp}(\vR)|^2 = \frac{1}{2\epsilon_0} \int d\vR \int d\vR' \hP(\vR)  \overleftrightarrow{\vdelta}_{\perp}(\vR - \vR')  \hP(\vR')$.
        }   
        \begin{equation}\label{eq:H_int_normal_mode}
            \begin{aligned}
            \hH_{\rm int} = &- \sum_{k\lambda}\sum_{n=1}^{N} 
            iD_{k}
            \left [ \hat{a}_{k\lambda} \vf_{k\lambda}(\vR_n)  - \hat{a}^{\dagger}_{k\lambda} \vf_{k\lambda}^{\ast}(\vR_n) \right ] \cdot \hat{\vmu}^{(n)}  \\ 
            &+ \sum_{k\lambda} \frac{1}{2\epsilon_0}\left | \sum_{n=1}^{N} \hat{\vmu}^{(n)}\cdot \vf_{k\lambda}(\vR_n)\right |^2 .
            \end{aligned}
        \end{equation}

        To proceed, we further rewrite $\hat{a}_{k\lambda}$ and $\hat{a}_{k\lambda}^{\dagger}$ as $\hat{p}_{k\lambda}$ and $\hat{q}_{k\lambda}$:
        \begin{subequations}\label{eq:a_adagger}
            \begin{align}
                \hat{a}_{k\lambda} &= \frac{1}{\sqrt{2\hbar\omega_{k}}}(\omega_{k} \hat{q}_{k\lambda} + i \hat{p}_{k\lambda}) , \\
                \hat{a}_{k\lambda}^{\dagger} &= \frac{1}{\sqrt{2\hbar\omega_{k}}}(\omega_{k} \hat{q}_{k\lambda} - i \hat{p}_{k\lambda})  .
            \end{align}
        \end{subequations}
        When Eq. \eqref{eq:a_adagger} is substituted  into $\hH_{\rm ph}$ in Eq. \eqref{eq:H_ph_second_quant} as well as $\hH_{\rm int}$ in Eq. \eqref{eq:H_int_normal_mode},  the field-related Hamiltonian $\hH_{\rm F} \equiv \hH_{\rm ph} + \hH_{\rm int}$ can be expressed as:
        \begin{equation}\label{eq:HF_before_exchange}
        \begin{aligned}
            \hH_{\rm F} &= \sum_{k\lambda} \frac{1}{2}\omega_{k}^2 \left( \hat{q}_{k\lambda} - \sum_{n=1}^{N} \frac{\hat{\vmu}^{(n)}}{\sqrt{\epsilon_0}\omega_{k}} \cdot \text{Im}\left[ \vf_{k\lambda}(\vR_n) \right] \right)^2 \\
            & + \frac{1}{2}\left( \hat{p}_{k\lambda}  - \sum_{n=1}^{N} \frac{\hat{\vmu}^{(n)}}{\sqrt{\epsilon_0}}\cdot \text{Re}\left[ \vf_{k\lambda}(\vR_n) \right] \right)^2  .
        \end{aligned}
        \end{equation}
        At this point, we assume that $\vf_{k\lambda}(\vR_n)$ is always real-valued, so $\text{Im}\left[ \vf_{k}(\vR_n) \right] \equiv 0$ and only $\hat{p}_{k\lambda}$ is coupled to the molecules. Because photons are harmonic oscillators, we may also exchange  $\hat{p}_{k\lambda}$ and $\hat{q}_{k\lambda}$ in Eq. \eqref{eq:HF_before_exchange}. Practically speaking, we may redefine $\hat{p}_{k\lambda}'=\omega_{k}\hat{q}_{k\lambda}  $ and $\hat{q}_{k\lambda}'=-\frac{1}{\omega_{k}}\hat{p}_{k\lambda}  $. After this exchange, the canonical commutation relation is preserved: $ [\hat{p}_{k\lambda}', \hat{q}_{k\lambda}'] = [\hat{p}_{k\lambda}, \hat{q}_{k\lambda}] = i\hbar$, and the Hamiltonian becomes:
        \begin{equation}\label{eq:H_F_updated}
            \hH_{\rm F} = \sum_{k\lambda} \frac{1}{2}\hat{p}_{k\lambda}^2  + \frac{1}{2}\omega_{k}^2 \left( \hat{q}_{k\lambda} + \sum_{n=1}^{N} \frac{\hat{\vmu}^{(n)} \cdot  \vf_{k\lambda}(\vR_n) }{\sqrt{\epsilon_0}\omega_{k}}  \right)^2 .
        \end{equation}
        Eq. \eqref{eq:H_F_updated} can be regarded as a generalized, multi-mode  form of the widely used  single-mode Pauli--Fierz  Hamiltonian.
        
        \subsubsection{Toward molecular dynamics simulations}
        
        To be suitable for molecular dynamics simulations, the quantum position and momentum operators of photons are rescaled as
        \begin{subequations}
            \begin{align}
            \hat{p}_{k\lambda}
                & =  \hat{\widetilde{p}}_{k\lambda}/ \sqrt{m_{k\lambda}} , \\
             \hat{q}_{k\lambda} 
                & =  \hat{\widetilde{q}}_{k\lambda} \sqrt{m_{k\lambda}} ,
            \end{align}
        \end{subequations}
        where $m_{k\lambda}$ denotes the auxiliary mass of the photon mode, which is introduced to make the photonic dynamics compatible with conventional molecular dynamics packages.
        With this rescaling, the field-related Hamiltonian in Eq. \eqref{eq:H_F_updated} becomes
        \begin{equation}\label{eq:HF_rescaled}
        \begin{aligned}
            \hH_{\rm F} &= \sum_{k\lambda} \frac{1}{2m_{k\lambda}}\hat{\widetilde{p}}_{k\lambda}^2  \\ &+ \frac{1}{2}m_{k\lambda}\omega_{k}^2 \left( \hat{\widetilde{q}}_{k\lambda} + \sum_{n=1}^{N} \frac{\hat{\vmu}^{(n)} \cdot  \vf_{k\lambda}(\vR_n) }{\sqrt{\epsilon_0 m_{k\lambda}}\omega_{k}}  \right)^2 .
        \end{aligned}
        \end{equation}
        Finally, we project the QED Hamiltonian in Eq. \eqref{eq:HF_rescaled} to the electronic ground state. Under this cavity Born--Oppenheimer approximation, \cite{Flick2017cBO} the light-matter Hamiltonian becomes
        \begin{subequations}\label{eq:H_ground_final}
        \begin{equation}
            \hH_{\rm QED}^{\rm G} = \hH_{\rm M}^{\rm G} + \hH_{\rm F}^{\rm G} ,
        \end{equation}
        where $\hH_{\rm M}^{\rm G}$ denotes the conventional (kinetic + potential) molecular Hamiltonian in the electronic ground state, and the field-related Hamiltonian becomes
            \begin{equation}
            \begin{aligned}
             \hH_{\rm F}^{\rm G} &= \sum_{k\lambda} \frac{1}{2m_{k\lambda}}\hat{\widetilde{p}}_{k\lambda}^2  \\
             & + \frac{1}{2}m_{k\lambda}\omega_{k}^2 \left( \hat{\widetilde{q}}_{k\lambda} + \sum_{n=1}^{N} \frac{\hat{\vmu}_{\rm g}^{(n)}\cdot  \vf_{k\lambda}(\vR_n)}{\sqrt{\epsilon_0 m_{k\lambda}}\omega_{k}}  \right)^2 .
             \end{aligned}
            \end{equation}
            Here, $\hat{\vmu}_{\rm g}^{(n)}$ denotes the dipole operator of the $n$-th molecule under electronic ground state.
        \end{subequations}
        
        The electronic ground-state light-matter Hamiltonian in Eq. \eqref{eq:H_ground_final} resembles the Hamiltonian in Ref. \citenum{Li2020Water}. The only difference is that here the long wave approximation is not used and the spatial dependence of the photonic mode functions  is explicitly considered. If the long wave approximation is taken here, $\vf_{k\lambda}(\vR_n) \equiv \vxi_{\lambda}/\sqrt{\Omega}$ (i.e., there is no spatial variance in the photonic mode functions), Eq. \eqref{eq:H_ground_final}  is reduced to the light-matter Hamiltonian in Ref. \citenum{Li2020Water}.

        \subsection{Efficient mesoscale CavMD simulations}

        After obtaining the generalized light-matter Hamiltonian in Eq. \eqref{eq:H_ground_final}, we are interested in deriving an efficient simulation scheme of CavMD. 
        
        \subsubsection{A brute-force approach}
        Given the quantum Hamiltonian in Eq. \eqref{eq:H_ground_final}, the classical equations of motion of the coupled light-matter system become
        \begin{subequations}\label{eq:EOM_original}
            \begin{align}
                M_{\alpha} \ddot{\mathbf{R}}_{\alpha} &= \mathbf{F}_{\alpha}^{(0)} + \mathbf{F}_{\alpha}^{\rm cav} , \\
                m_{k\lambda} \ddot{\widetilde{q}}_{k\lambda} &= - m_{k\lambda} \omega_{k}^2\widetilde{q}_{k\lambda} - \varepsilon_{k\lambda} d_{\text{g},k\lambda} .
            \end{align}
        Here, $d_{\text{g},k\lambda}$ denotes the total molecular dipole moment projected along the cavity polarization direction:
        \begin{equation}\label{eq:dipole_along_mode}
            d_{\text{g},k\lambda} \equiv \sum_{n=1}^{N} \sqrt{\mathcal{V}}\vmu_{\rm g}^{(n)}\cdot \vf_{k\lambda}(\vR_n) ,
        \end{equation}
        \end{subequations}
        where $\mathcal{V}$, in the units of volume, denotes the normalization coefficient of the mode function which satisfies $\avg{ \mathcal{V} |\vf_{k\lambda}(\vR)|^2} = 1$. This normalization guarantees $\sqrt{\mathcal{V}} \vf_{k\lambda}(\vR_n)$ to be dimensionless, so $d_{\text{g},k\lambda}$ maintains the same units as $\vmu_{\rm g}^{(n)}$.  $M_{\alpha}$ and $\mathbf{R}_{\alpha}$ denote the mass and position of the $\alpha$-th nucleus, $\mathbf{F}_{\alpha}^{(0)}$ denotes the nuclear force acting on the $\alpha$-th nucleus without the cavity (i.e., the same force as that in conventional molecular dynamics simulations), and $\mathbf{F}_{\alpha}^{\rm cav}$ denotes the  force from the cavity acting on the the $\alpha$-th nucleus: 
            \begin{equation}\label{eq:Fcav_alpha}
                \mathbf{F}_{\alpha}^{\rm cav} = -\sum_{k\lambda} \left(
                \varepsilon_{k\lambda} \widetilde{q}_{k\lambda} + 
                \frac{\varepsilon_{k\lambda}^2}{m_{k\lambda} \omega_{k}^2} d_{\text{g},k\lambda}
                \right) \frac{\partial d_{\text{g},k\lambda}}{\partial \mathbf{R}_{\alpha}} ,
            \end{equation}
            where the effective light-matter coupling  $\varepsilon_{k\lambda}$ is defined as
        \begin{equation}
            \varepsilon_{k\lambda} \equiv \sqrt{\frac{m_{k\lambda}\omega_{k}^2}{\mathcal{V} \epsilon_0} } .
        \end{equation}
        Because the dielectric boundary conditions determine the photonic mode functions, given an explicit form of $\vf_{k\lambda}(\vR_n)$,  Eq. \eqref{eq:EOM_original}  allows for directly propagating the coupled photon-molecular dynamics in the dielectric environment. However, due to the large number of molecules and the infinite number of photonic modes involved in planar Fabry--P\'erot cavities, a brute-force propagation of Eq. \eqref{eq:EOM_original} is unrealistic. 
        
        Below, using a rectangular parallelepiped as the photonic boundary conditions, we demonstrate how to efficiently simulate CavMD in the mesoscale with an affordable computational cost.

        \subsubsection{Photonic mode functions in a rectangular parallelepiped}

        \begin{figure}
		\centering
		\includegraphics[width=0.6\linewidth]{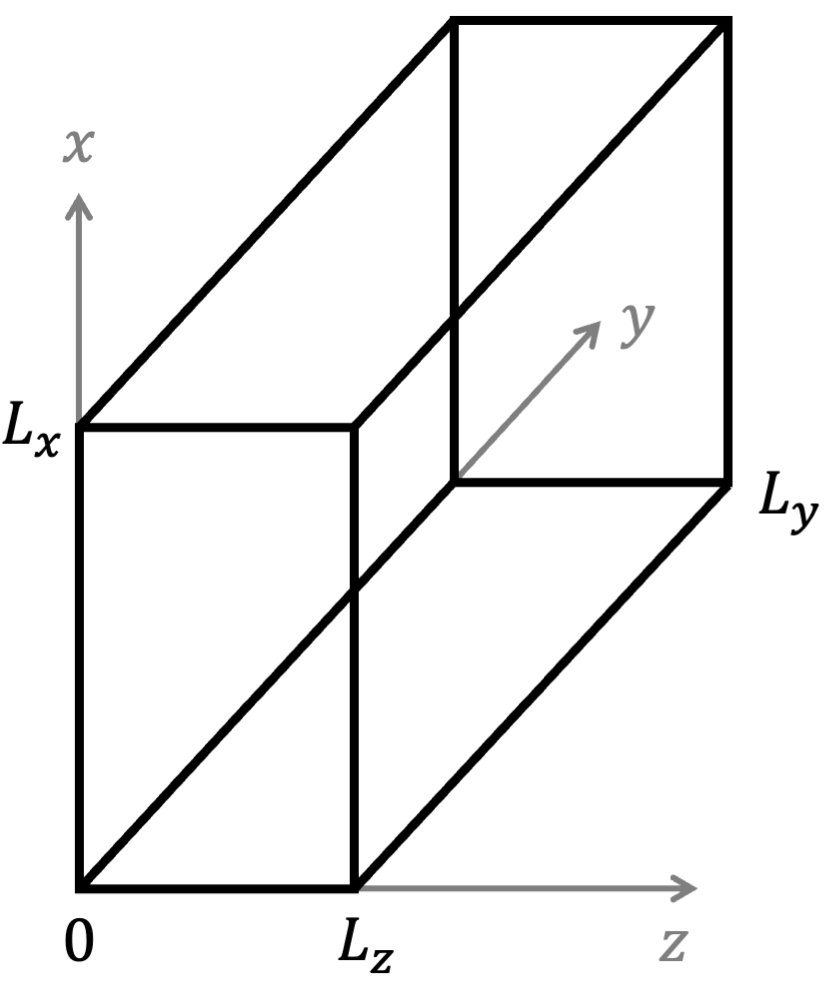}
		\caption{
        An optical resonator with volume $\Omega$ shaped as a rectangular parallelepiped, with lengths along the three dimensions denoted as $L_x$, $L_y$, and $L_z$. The electric field amplitude is assumed to be zero at the boundaries.
		}
		\label{fig:box}
        \end{figure}
        
        For a rectangular parallelepiped shown in Fig. \ref{fig:box}, the photonic mode functions are \cite{Schleich2001}
        \begin{subequations}
            \begin{align}
                f_{k,x} & = \sqrt{\frac{8}{\Omega}} \cos(k_x x) \sin(k_y y) \sin(k_z z), \\
                f_{k,y} & = \sqrt{\frac{8}{\Omega}} \sin(k_x x) \cos(k_y y) \sin(k_z z), \\
                f_{k,z} & = \sqrt{\frac{8}{\Omega}} \sin(k_x x) \sin(k_y y) \cos(k_z z). 
            \end{align}
        \end{subequations}
        Here, $\Omega$ denotes the volume of the rectangular parallelepiped;
        \begin{subequations}
            \begin{align}
                k_x &= \frac{l_x\pi}{L_x}, \\
                k_y &= \frac{l_y\pi}{L_y}, \\
                k_z &= \frac{l_z\pi}{L_z}, 
            \end{align}
        \end{subequations}
        where $L_x$, $L_y$,  and $L_z$ denote the lengths of the rectangular parallelepiped in three dimensions, and $l_x$, $l_y$, and $l_z$ are independent integers.

\begin{figure*}
		\centering
		\includegraphics[width=0.8\linewidth]{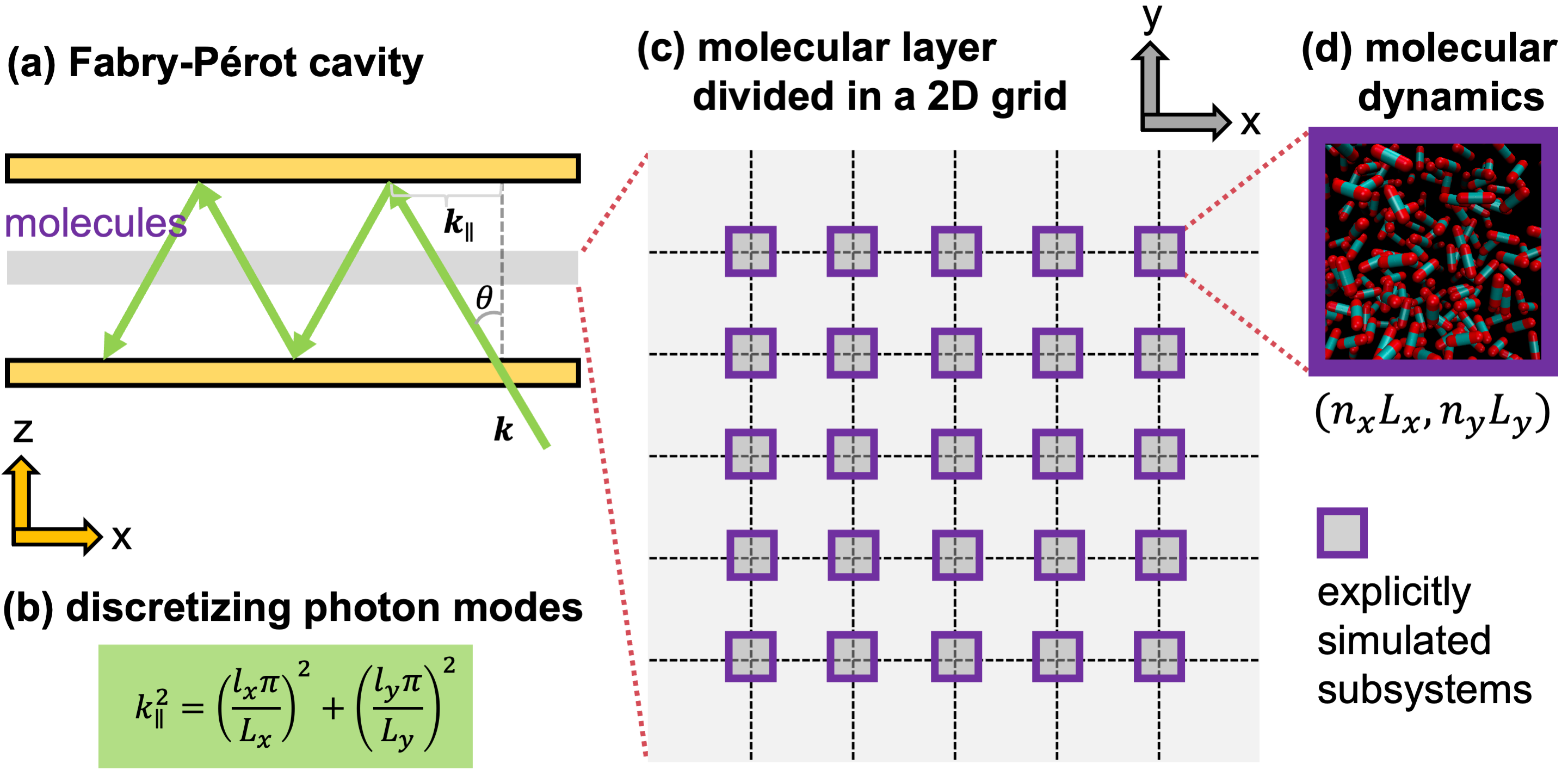}
		\caption{Strategy for simulating Fabry--P\'erot vibraitonal strong coupling. (a) The Fabry--P\'erot cavity viewed in the $xz$ plane.
        A  layer of molecules is confined at the middle of the cavity. (b) A finite number of photon modes are simulated by discretizing $\vk_{\parallel}$. (c) The molecules are divided in a 2D grid along the $xy$ plane. (d) For an efficient simulation of the macroscopic molecular system, only the small molecular subsystems at different grid points (labeled as the purple squares) are explicitly simulated. 
		}
		\label{fig:molecular_grid}
	\end{figure*}

        This rectangular parallelepiped can be used to simulate CavMD dynamics in planar Fabry--P\'erot cavities. To this end, we  assume the cavity is placed along the $z$ direction and $L_z \ll L_x, L_y$. If only the fundamental cavity mode of $k_z$ is relevant to our research interest (e.g., when the high-frequency vibrational mode of molecules is near resonant with the fundamental cavity mode), we can further take $k_z = \pi/L_z$ and ignore any higher order modes. If the molecules are placed near the plane of $z = L_z/2$ (as demonstrated in Fig. \ref{fig:molecular_grid}a), only the following mode functions have a non-zero contribution:
        \begin{subequations}\label{eq:mode_function_2d}
            \begin{align}
                f_{k,x} & = \sqrt{\frac{8}{\Omega}} \cos(k_x x) \sin(k_y y) ,\\
                f_{k,y} & = \sqrt{\frac{8}{\Omega}} \sin(k_x x) \cos(k_y y) .
            \end{align}
        \end{subequations}
        For this simplified cavity setup, the normalization coefficient $\mathcal{V} = \Omega/2$, so $\avg{\mathcal{V}|f_{k,x}|^2} = \avg{\mathcal{V}|f_{k,y}|^2}=1$. The corresponding cavity mode frequency $\omega_{k}$ becomes
        \begin{equation}
            \omega_{k} = c \sqrt{\left(\frac{\pi}{L_z}\right )^2 + k_{\parallel}^2 } .
        \end{equation}
        Here, $c = c_0/n_{\rm ref}$ denotes the speed of light in the medium; $c_0$ denotes the speed of light in the vacuum and $n_{\rm ref}$ denotes the refractive index of the medium. The in-plane wave vector $k_{\parallel}$ is defined as
        \begin{equation}\label{eq:k_parallel_2d}
            k_{\parallel}^2 = \left(\frac{l_x\pi}{L_x}\right )^2 + \left(\frac{l_y\pi}{L_y}\right )^2 .
        \end{equation}
        In practice, we include a finite number of cavity modes with $1 \leq l_x \leq l_x^{\rm max}$ and $1 \leq l_y \leq l_y^{\rm max}$.

        \subsubsection{A 2D grid of molecules}
        
        Although we have assumed that  the molecules of interest are confined to a thin layer along the plane $z=L_z/2$ (Fig. \ref{fig:molecular_grid}a), directly simulating all $N$  molecules in this layer is computationally  expensive.  Thus, we approximate  the full molecular system  by  a smaller number of grid points (Fig. \ref{fig:molecular_grid}c), where each grid point represents a finite-sized,  explicitly simulated molecular system. This grid representation  allows us to simulate the system with bearable computational resources.

        With the 2D grid representation, the total molecular dipole $d_{\text{g},k\lambda}$ defined in Eq. \eqref{eq:dipole_along_mode} can be calculated as 
        \begin{equation}\label{eq:dipole_tot_grid_rep}
            d_{\text{g},k\lambda} = N_{\rm ratio} d_{\text{g},k\lambda}^{\rm sub} .
        \end{equation}
        Here, $N_{\rm ratio}$ denotes the ratio between the true number of molecules in the cavity versus the number of explicitly simulated molecules; $d_{\text{g},k\lambda}^{\rm sub}$ denotes the total dipole moment of the explicitly simulated molecules (the purple areas in Fig. \ref{fig:molecular_grid}c),
        \begin{equation}\label{eq:dipole_sub}
            d_{\text{g},k\lambda}^{\rm sub} = \sum_{\vec{n}=(n_x,n_y)} \sqrt{\mathcal{V}}\vmu_{\rm g}^{(\vec{n})}\cdot \vf_{k\lambda}(\vR_{\vec{n}}) ,
        \end{equation}
        where $\vmu_{\rm g}^{(\vec{n})}$ denotes the total molecular dipole in each explicitly simulated grid point, and $\vR_{\vec{n}}$ denotes the position of each grid point labeled by $\vec{n} = (n_x, n_y)$.

        \subsubsection{A practical classical simulation scheme}
        
        With the molecular dipole defined above, we can rewrite the equations of motion in Eq. \eqref{eq:EOM_original} as
        \begin{subequations}\label{eq:EOM_new}
            \begin{align}
                M_{\alpha} \ddot{\mathbf{R}}_{\alpha} &= \mathbf{F}_{\alpha}^{(0)} + \mathbf{F}_{\alpha}^{\rm cav} , \\
                m_{k\lambda} \ddot{\dbtilde{q}}_{k\lambda} &= - m_{k\lambda} \omega_{k}^2\dbtilde{q}_{k\lambda} - \widetilde{\varepsilon}_{k\lambda} d_{\text{g},k\lambda}^{\rm sub} , \label{eq:EOM_new_ph}
            \end{align}
            where the cavity force is 
            \begin{equation}
                \mathbf{F}_{\alpha}^{\text{cav}} = -\sum_{k\lambda} \left(
                \widetilde{\varepsilon}_{k\lambda} \dbtilde{q}_{k\lambda} + 
                \frac{ \widetilde{\varepsilon}_{k\lambda}^2}{m_{k\lambda} \omega_{k}^2} d_{\text{g},k\lambda}^{\rm sub}
                \right) \frac{\partial d_{\text{g},k\lambda}^{\rm sub} }{\partial \mathbf{R}_{\alpha}} .
            \end{equation}
        \end{subequations}
        Here, we have defined the rescaled photon position $\dbtilde{q}_{k\lambda}$ as
        \begin{equation}\label{eq:q_dbtilde}
            \dbtilde{q}_{k\lambda} \equiv \widetilde{q}_{k\lambda} / \sqrt{N_{\rm ratio}}
        \end{equation}
        and the new effective light-matter coupling as
        \begin{equation}\label{eq:varepslion_tilde}
            \widetilde{\varepsilon}_{k\lambda} = \sqrt{N_{\rm ratio}} \varepsilon_{k\lambda}  = \sqrt{\frac{N_{\rm ratio}m_{k\lambda}\omega_{k}^2}{\mathcal{V} \epsilon_0} } .
        \end{equation}
        The rescaling introduced in Eqs. \eqref{eq:q_dbtilde} and \eqref{eq:varepslion_tilde} is necessary because it guarantees that the equations of motion in Eq. \eqref{eq:EOM_new} are of the same form as Eq. \eqref{eq:EOM_original}, except that (i) the total dipole moment of the explicitly simulated molecules $d_{\text{g},k\lambda}^{\rm sub}$ replaces the true total dipole moment of the whole molecular layer and (ii) $\widetilde{\varepsilon}_{k\lambda}$ becomes unphysically large. Of course, as demonstrated in previous CavMD work \cite{Li2020Nonlinear,Li2021Solute,Li2021Collective}, because of the potential for artifacts introduced by this rescaling procedure, it is necessary to validate whether the simulated cavity effects can persist in real experiments containing a macroscopic  number of molecules. This validation process can be done by increasing the number of explicitly simulated molecules as well as reducing $\widetilde{\varepsilon}_{k\lambda}$ (these two parameters should be changed in a manner such that the observed Rabi splitting remains the same) until the convergence of the simulation results.

        Eq. \eqref{eq:EOM_new} illustrates that three molecular quantities, i.e., $\mathbf{F}_{\alpha}^{(0)}$, $d_{\text{g},k\lambda}^{\rm sub}$, and $\partial d_{\text{g},k\lambda}^{\rm sub} / \partial \mathbf{R}_{\alpha}$, must be determined during CavMD simulations. Previously, under the single-mode limit, these three quantities have been evaluated using both non-polarizable empirical force fields \cite{Li2020Water} and hybrid QM/MM calculations \cite{Li2023QMMM}. 

        When non-polarizable force fields are used to describe the molecules, the total dipole moment of the molecular  system at each grid point $\vec{n}=(n_x, n_y)$ becomes
        \begin{equation}\label{eq:mu_each_grid_mm}
            \vmu_{\rm g}^{(\vec{n})} = \sum_{\alpha \in \vec{n}} Q_{\alpha} \mathbf{R}_{\alpha} ,
        \end{equation}
        where $Q_\alpha$ denotes the partial charge of the $\alpha$-th nucleus. 
        Therefore, according to Eq. \eqref{eq:dipole_sub}, $d_{\text{g},k\lambda}^{\rm sub}$ can be expressed as
        \begin{equation}\label{eq:dipole_MM}
            d_{\text{g},k\lambda}^{\rm sub} = \sum_{\vec{n}} \sum_{\alpha\in \vec{n}}\sqrt{\mathcal{V}} Q_{\alpha} \mathbf{R}_{\alpha}\cdot  \vf_{k\lambda}(\vR_{\vec{n}}) .
        \end{equation}
        Similarly, the dipole derivatives become
        \begin{equation}\label{eq:dipole_deriv_MM}
            \frac{\partial d_{\text{g},k\lambda}^{\rm sub}}{\partial R_{\alpha i}^{(\vec{n})} } =  \sqrt{\mathcal{V}} Q_{\alpha}  \ve_i\cdot \vf_{k\lambda}(\vR_{\vec{n}}) ,
        \end{equation}
        where $R_{\alpha i}^{(\vec{n})}$ represents the nuclear position along the $i$-th direction at grid point $\vec{n} = (n_x, n_y)$, and $\ve_i$ denotes the unit vector along the $i$-th direction. When the equations for the dipole and its derivatives [Eqs. \eqref{eq:dipole_MM} and \eqref{eq:dipole_deriv_MM}]  are substituted into the equations of motion [Eq. \eqref{eq:EOM_new}], the mesoscale cavity dynamics can be efficiently simulated. 
        
        Recent studies suggest that the dipole self-energy term in the light-matter Hamiltonian can be important and may induce additional intermolecular interactions.\cite{Horak2024,Schnappinger2023,Sidler2024} As a dynamics approach, CavMD includes the dipole self-energy term and evaluates its gradients on-the-fly, which contain dipole moments and their derivatives.  Of course, because dipole moments and their derivatives are important, a more accurate description of strong coupling may require a dipole surface beyond  the use of fixed point charges, e.g., using inputs from on-the-fly electronic structure theory or machine-learning dipole surfaces. Nevertheless, previous work has demonstrated that using fixed point charges to represent the dipole surface can yield polariton relaxation dynamics in qualitative agreement with experiments.\cite{Li2020Nonlinear,Li2021Relaxation} 

        As a side note, the current formalism of CavMD can be directly applied to study systems with well-defined, charge-neutral molecules. For this type of the systems, the dipole moment per grid point, $\vmu_{\rm g}^{(\vec{n})}$, is independent of periodic boundary conditions (applied in each molecular subsystem)  and the origin of the coordinate system. In other words, $\vmu_{\rm g}^{(\vec{n})}$ for well-defined, charge-neutral molecules is translational invariant. However, when the molecular subsystems contain free ions, extra care is needed when performing CavMD simulations. Because $\vmu_{\rm g}^{(\vec{n})}$ for free ions is translational variant (depending on periodic boundary conditions and the origin of the coordinate system), the translational motion of free ions may potentially contaminate the photonic dynamics through the light-matter coupling in Eq. \eqref{eq:H_F_updated}. One possible solution is to remove the free ion contribution when evaluating $\vmu_{\rm g}^{(\vec{n})}$ in Eq. \eqref{eq:mu_each_grid_mm}, thus ensuring the translationally invariant dipole moments. More extensive study is needed to explore the proper CavMD scheme for simulating molecular systems with free ions.

        \subsubsection{An outline of mesoscale CavMD simulation}

        Here, we give a brief outline on how to perform mesoscale CavMD simulations from scratch.
        \begin{itemize}
            \item[(i)] Determine the cavity mode structure by setting the fundamental cavity frequency $\omega_\perp = c \pi/L_z$,  cavity mirror plane lengths $L_x$, $L_y$, as well as the maximal mode numbers  $l_x^{\rm max}$, $l_y^{\rm max}$ along the cavity mirror plane [Eqs. \eqref{eq:mode_function_2d}-\eqref{eq:k_parallel_2d}].
            
            \item[(ii)] Determine the number and positions of molecular grid points along the $(x, y)$ plane, and setup the initial molecular geometry for each grid point [Eqs. \eqref{eq:mu_each_grid_mm}-\eqref{eq:dipole_deriv_MM}].
            
            \item[(iii)] Given the effective light-matter coupling $\widetilde{\varepsilon}_{k\lambda}$, run the mesoscale CavMD simulation using Eq. \eqref{eq:EOM_new}.
            
            \item[(iv)] (Optional) Increase the explicitly simulated molecular number $N_{\rm simu}$ (which depends on both $N_{\rm grid}$ and the number of molecules in each grid point) while reducing the effective coupling strength $\widetilde{\varepsilon}_{k\lambda} \propto 1/\sqrt{N_{\rm simu}}$ accordingly, and then check whether the observed cavity effect is robust for different cavity sizes. 
            The large $N_{\rm simu}$ limit may correspond to VSC in realistic planar Fabry--P\'erot cavities. 
        \end{itemize}
       
        \subsubsection{Connecting to previous CavMD work with a single cavity mode}

        The present formalism can be reduced to previous CavMD work with only a single cavity mode, which contains two degenerate cavity photons polarized along either the  $x$- or the $y$-direction. \cite{Li2020Water} This reduction is guaranteed because the equations of motion derived in Eq. \eqref{eq:EOM_new} are identical to those of our earlier work.\cite{Li2020Water,Li2020Nonlinear} 
                
        In detail, to include only two degenerate cavity photons which are polarized along either the $x$ or the $y$ direction, we set $k_x = \pi/L_x$, $k_y = \pi/L_y$, and $L_x = L_y \gg L_z$. Then, we may use only a single grid point to represent the whole molecular system, where the location of the grid point is denoted $x = \frac{L_x}{4}$ and $y = \frac{L_y}{4}$. With these parameters, both $k_x x$ and $k_y y$ equal $\pi/4$; according to Eq. \eqref{eq:mode_function_2d},  the two possible photon mode functions become
        \begin{subequations}
            \begin{align}
                f_{k,x} & = \sqrt{\frac{2}{\Omega}} = \sqrt{\frac{1}{\mathcal{V}}}, \\
                f_{k,y} & = \sqrt{\frac{2}{\Omega}} = \sqrt{\frac{1}{\mathcal{V}}}. 
            \end{align}
        \end{subequations}
        With the above parameters, the molecular dipole and its derivatives  [Eqs. \eqref{eq:dipole_MM} and \eqref{eq:dipole_deriv_MM}] become 
        \begin{subequations}
            \begin{equation}
            d_{\text{g},k\lambda}^{\rm sub} =  \sum_{\alpha } Q_{\alpha} \mathbf{R}_{\alpha}\cdot  \vxi_\lambda ,
        \end{equation}
        and
        \begin{equation}
            \frac{\partial d_{\text{g},k\lambda}^{\rm sub}}{\partial R_{\alpha i}} =  Q_{\alpha}  \ve_i\cdot \vxi_\lambda ,
        \end{equation}
        \end{subequations}
        which are identical to previous work. \cite{Li2020Water} Hence, the current implementation is consistent with previous CavMD work with a single cavity mode.

        \subsubsection{Path-integral simulations of nuclear and photonic quantum effects}

        Before ending this section, we briefly discuss how to include quantum effects in mesoscale CavMD simulations. Although the above equations of motion in Eq. \eqref{eq:EOM_new} are classical, nuclear and photonic quantum effects can also be captured by performing path-integral simulations. \cite{Li2022RPMDCav,Lieberherr2023CavMD} After all,  in  Eq. \eqref{eq:EOM_new}, photons behave just like additional    ``nuclei''. Hence, one can directly apply different flavors of path-integral approaches developed over the past decades \cite{Markland2018} to study VSC.

       \begin{figure*}
		\centering
		\includegraphics[width=0.8\linewidth]{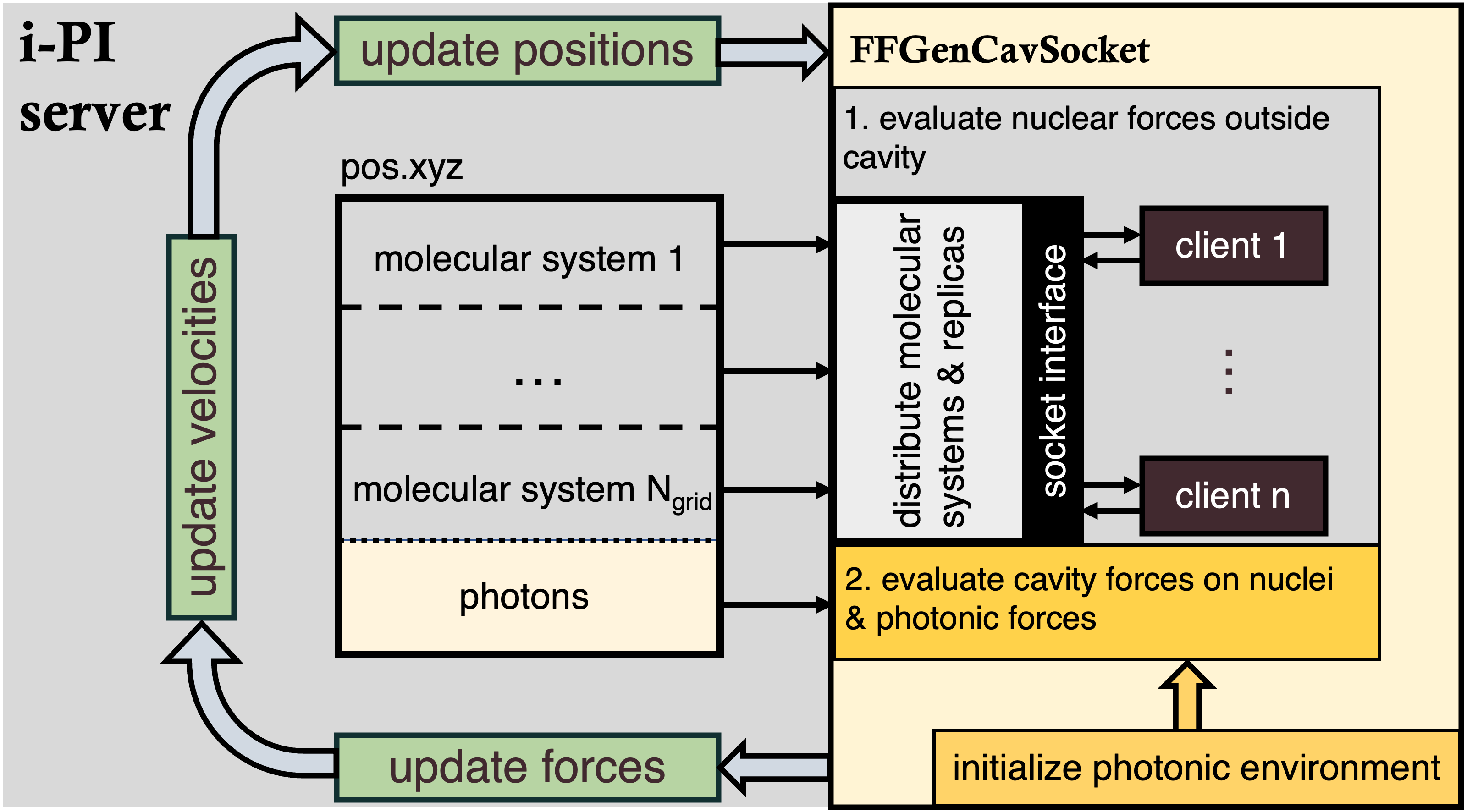}
		\caption{A sketch of the mesoscale CavMD scheme as implemented in i-PI.  The original i-PI server is responsible for updating the positions and velocities of the entire molecular + photonic system. The newly implemented \texttt{FFGenCavSocket} method, which stands for the \texttt{f}orce \texttt{f}ield \texttt{socket} interface for a \texttt{gen}eral \texttt{cav}ity setup, evaluates the forces for the whole system, including both the nuclear forces at different grid points (as shown in Fig. \ref{fig:molecular_grid}c) and the photonic forces. In path-integral simulations with $P$ beads, the \texttt{FFGenCavSocket} method can connect to $N_{\rm grid} \times P$ different clients in parallel, thus significantly reducing the gradient evaluation time.
        }
		\label{fig:i-pi-implementation}
    \end{figure*}
    
    \section{Simulation details}\label{sec:simulation_details}

    We implemented the mesoscale CavMD scheme in the open source package i-PI; \cite{Kapil2019} see the caption of Fig. \ref{fig:i-pi-implementation} for a detailed description of the implementation. In particular, the forces of molecular systems at different grid points were evaluated in parallel by i-PI through the socket interface by connecting to multiple clients; each client runs its own force evaluation code. The code and input files are available on Github at \url{https://github.com/TaoELi/cavity-md-ipi}.

\begin{figure*}
		\centering
		\includegraphics[width=1.0\linewidth]{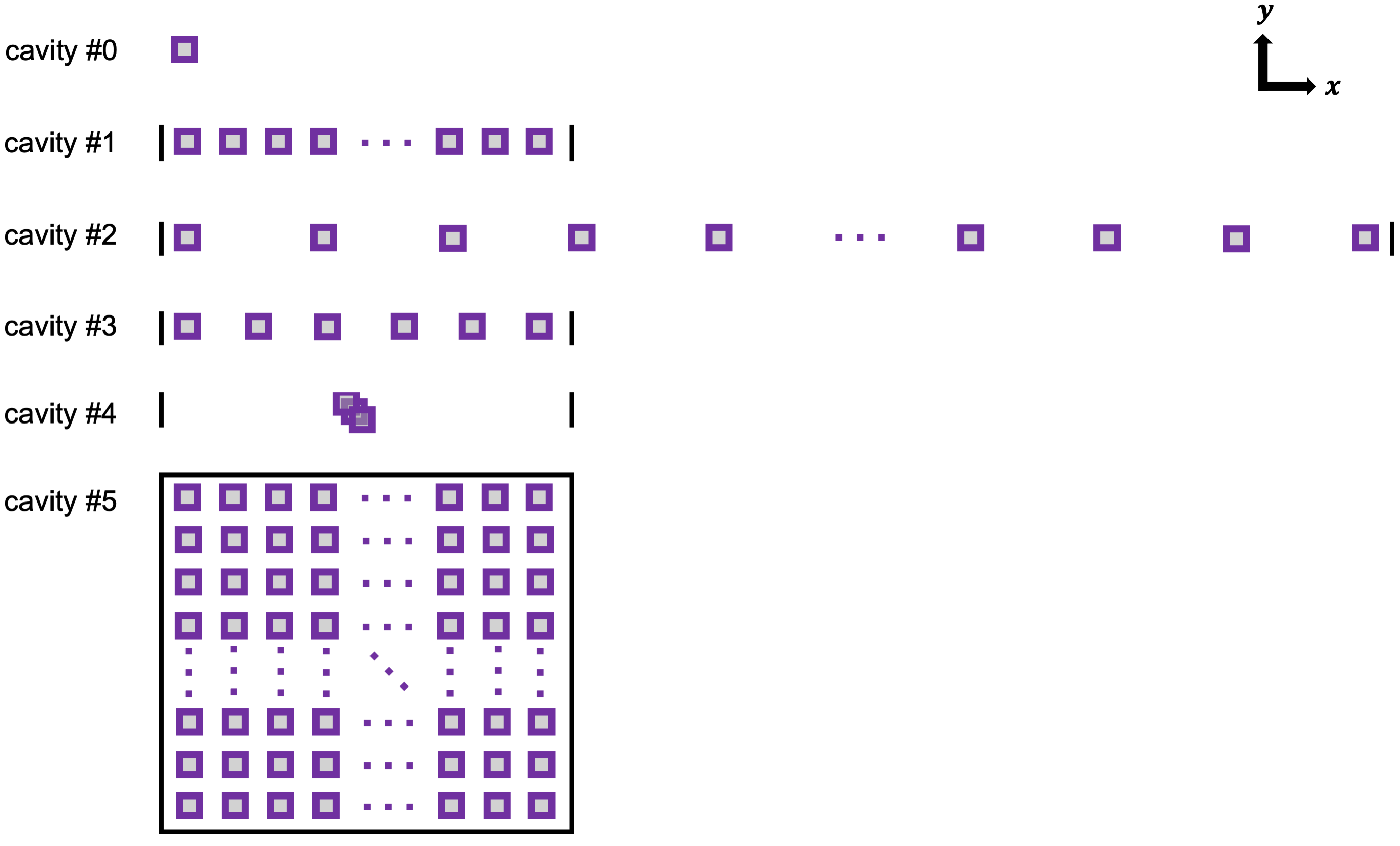}
		\caption{Sketch of different cavity setups. Cavity \#0 represents the standard cavity setup in previous CavMD simulations, where a large ensemble of molecules is coupled to a single cavity mode. Cavity \#1-\#5 represent the five cavity setups simulated in this work; see Table \ref{table:cavity_parameters} for the corresponding cavity parameters.
		}
		\label{fig:cavity_setup}
    \end{figure*}

    The implementation of our method was tested on various cavity setups.
    The sketch of these cavity setups and the corresponding parameters are shown in Fig. \eqref{fig:cavity_setup} and Table \ref{table:cavity_parameters}. In each cavity setup, the whole molecular system was represented by $N_{\rm grid}$ grid points. In cavity \#1-\#4, because the grid points along the $y$-direction all located at the origin ($y=0$), as shown in Eq. \eqref{eq:mode_function_2d}, the mode functions along the $x$-direction vanished. In other words, in cavity \#1-\#4, cavity modes polarized along the $x$-direction were decoupled from the molecules. Throughout the paper, the cavity loss is set as zero. The method and results for including the cavity loss are briefly discussed in the Appendix. The auxiliary mass of the photon modes is set to $m_{\rm c} = m_{l\lambda} = 1$ a.u.

    \begin{table*}
    \caption{Parameters for different cavity setups. In the table, $\Delta \omega_{x,y} = c\pi/L_{x,y}$;  $\Delta \omega_{y} = 0$ indicates  $L_y = \infty$. Each array along the $x$- or $y$-direction represents the evenly spaced locations (in the units of $L_x$ or $L_y$) of the molecular grid points along the corresponding axis. For example, in cavity \#1, [0.027, 0.054, $\cdots$, 0.973] represent the $x$-grid  locations for $N_{\rm grid} = 36$ grid points. At each grid point, 36 \ch{CO2} molecules are explicitly simulated under periodic boundary conditions. Cavity \#1-\#4 represent four sets of parameters for 1D cavities, while cavity \#5 simulates a 2D cavity. The effective coupling strength $\widetilde{\varepsilon}$ refers to $\widetilde{\varepsilon}_{k\lambda}(\omega_{\perp})$, as $\widetilde{\varepsilon}_{k\lambda}$ is proportional to $\omega_{k}$ [Eq. \eqref{eq:varepslion_tilde}].  }
    \begin{adjustbox}{width=\linewidth,center}
    \label{table:cavity_parameters}
    \begin{tabular}{lccccccccr}
    \hline
    \hline
    cavity   & $\omega_{\perp}$ [cm$^{-1}$] \ & $\Delta\omega_{x}$ [cm$^{-1}$]\ & $\Delta \omega_{y}$ [cm$^{-1}$] \ & $l^{\rm max}_{x}$ \ & $l^{\rm max}_{y}$ \ & $N_{\rm grid}$ \ & $x$ grid positions [$L_x$] \ & $y$ grid positions [$L_y$] \ & $\widetilde{\varepsilon}$ [a.u.]\\
    \hline
    \#1   & 2320  & 50 & 0 & 36 & 1 & 36 & [0.027, 0.054, $\cdots$, 0.973] &  [0.0] & $5\times 10^{-5}$ \\
    \#2   & 2320  & 16.67 & 0 & 108 & 1 & 36 & [0.027, 0.054, $\cdots$, 0.973] &  [0.0] & $5\times 10^{-5}$ \\
    \#3   & 2320  & 50 & 0 & 36 & 1 & 6 & [0.143, 0.286, $\cdots$, 0.857] &  [0.0] & $1.225\times 10^{-4}$ \\
    \#4   & 2320  & 50 & 0 & 36 & 1 & 36 & [0.5, 0.5, $\cdots$, 0.5] &  [0.0] & $5\times 10^{-5}$ \\
    \#5   & 2320  & 50 & 50 & 18 & 18 & 1296 & [0.027, 0.054, $\cdots$, 0.973] &  [0.027, 0.054, $\cdots$, 0.973] & $8.3\times 10^{-6}$ \\
    \hline
    \hline
    \end{tabular}
    \end{adjustbox}
    \end{table*}

    Each grid point was assigned a periodic boundary condition box containing 36 \ch{CO2} molecules, which were explicitly simulated. The molecular density of the box was set to 1.101 g/cm$^3$. The \ch{CO2} molecules were modeled with the same force field that was used in previous work, which has been validated against experimental data.\cite{Li2020Nonlinear} Intermolecular Coulombic interactions were  calculated using Ewald summation.

    For equilibrium classical  simulations, the coupled cavity-molecular system was initially equilibrated for 20 ps under an NVT ensemble at 300 K, using a time step of 0.5 fs. A Langevin thermostat with a friction lifetime of 100 fs was attached to each particle (both photons and nuclei). After this equilibration simulation, 40 consecutive 20-ps NVE simulations were performed using a time step of 0.5 fs. At the beginning of each NVE simulation, the initial velocities of all particles were resampled with a Maxwell--Boltzmann distribution at 300 K. For nonequilibrium classical  simulation of polariton-polariton scattering in cavity \#1, given the 40 starting geometries of the above consecutive NVE simulation, the photonic energy of the $l_x = 12$ cavity mode was reset to 195 quanta. In practice, this initial excitation was implemented by setting the  $y$-coordinate of the $l_x = 12$ cavity mode to $y_{\rm c}=100$ \AA, according to $\frac{1}{2}m_{\rm c}\omega_c^2 y_{\rm c}^2 = 195\ \hbar\omega_{\rm c}$. With each of these perturbed geometries, a 20-ps NVE simulation was performed.

    For equilibrium path integral simulations, the thermostatted ring polymer molecular dynamics (TRPMD) approach was used. \cite{Rossi2014} While the simulation conditions of TRPMD largely resembled the classical case, here the difference is highlighted. We initialized the ring polymer with 4 beads for all nuclei and cavity photons with a time step of 0.25 fs. During all TRPMD simulations, a path integral Langevin equation  (PILE) thermostat \cite{Rossi2014} was  attached to the internal modes of the ring polymer with an optimally damped coefficient $\lambda=1/2$.

    Outside the cavity, the IR spectrum of liquid \ch{CO2} was calculated by  Fourier transforming the dipole autocorrelation function of the molecules:
    \cite{McQuarrie1976,Gaigeot2003,Habershon2008,Nitzan2006,Li2020Nonlinear}
	\begin{equation}\label{eq:IR_equation_molecule}
	\begin{aligned}
	F(\omega) \propto \omega^2 \int_{-\infty}^{+\infty} dt \ e^{-i\omega t}  \avg{\sum_{i=x,y,z}\mu_i(0)\mu_i(t)} .
	\end{aligned}
	\end{equation} 
    Here, $\mu_i(t)$ denotes the total dipole moment of the molecular system in each dimension. 

    Inside the cavity, the IR spectrum of the polaritons was calculated by evaluating the position autocorrelation function of the cavity photons:
    \begin{equation}\label{eq:IR_equation_photon}
	\begin{aligned}
	F_{k\lambda}(\omega) \propto \omega^2 \int_{-\infty}^{+\infty} dt \ e^{-i\omega t}  \avg{\dbtilde{q}_{k\lambda}(0)\dbtilde{q}_{k\lambda}(t)} .
	\end{aligned}
	\end{equation} 
    
    \section{Results}\label{sec:results}

    \begin{figure*}
		\centering
		\includegraphics[width=1.0\linewidth]{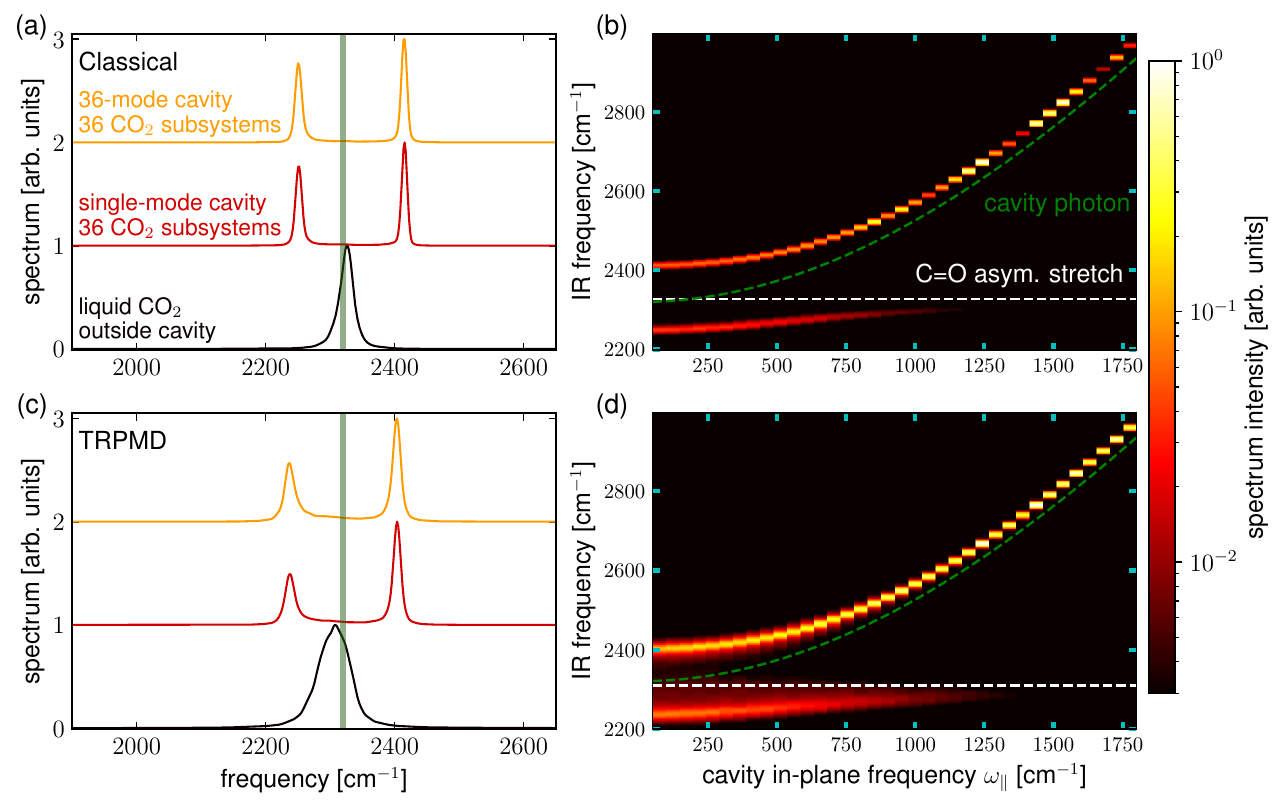}
		\caption{Polariton spectra in a lossless cavity from classical (upper panel) or path-integral (bottom panel) CavMD simulations. (a) The normalized IR spectrum of liquid \ch{CO2} outside the cavity (black), coupled to a single-mode cavity (red), or a multimode Fabry--P\'erot-like 1D cavity (cavity \#1, yellow). The vertical green line denotes the cavity frequency at normal incidence (the lowest-frequency cavity mode). (b) The polariton spectrum vs the cavity in-plane frequency $\omega_{\parallel}$ corresponding to the multimode cavity.  (c, d) The same plot as the upper panel except that the TRPMD method is used to propagate the dynamics.
		}
		\label{fig:IR_CO2}
    \end{figure*}

    \subsection{Cavity \#1: An 1D approximation of planar Fabry--P\'erot cavities}
    
    As shown in Fig. \ref{fig:cavity_setup}, cavity \#1 contains 36 cavity modes as well as 36 evenly spaced grid points along the $x$ direction.  For such a 1D cavity setup, the cavity frequency dispersion relation reads
    \begin{equation}
        \omega_{\text{c}}(l_x) = \sqrt{\omega_{\perp}^2 + l_x^2 \Delta \omega_{x}^2 } ,
    \end{equation}
    where $\omega_{\perp} = 2320$ cm$^{-1}$, $\Delta \omega_{x} = 50$ cm$^{-1}$, and $l_x = 1, 2, \cdots, 36$.

    \subsubsection{Polariton spectrum: the validity of the single-mode approximation}
        
    Fig. \ref{fig:IR_CO2}a shows the polariton spectrum (yellow line) 
    of \ch{CO2} confined in this cavity, obtained from classical CavMD simulations. The yellow line represents the polariton spectrum formed between the asymmetric \ch{C=O} stretch (black line) of liquid \ch{CO2} and the cavity photon mode at normal incidence (vertical green line).  This polariton spectrum is calculated by taking the Fourier transform of  the coordinate autocorrelation function of the lowest-frequency cavity mode (with $l_x = 1$). If only a single cavity mode at $l_x = 1$ is included in the simulation, the resulting polariton spectrum (red line) is identical to that of the 36-mode cavity \#1. This agreement shows the validity of using the single-mode approximation for describing the linear-response polariton spectrum:  Cavity modes at different $\omega_{\parallel}$ values effectively do not interact with each other. Of course, for cavity \#1, when the coordinate autocorrelation functions of cavity modes at different $\omega_{\parallel}$ are evaluated, the well-known $\omega_{\parallel}$-dependent UP and LP branches  \cite{Li2022Review,Simpkins2023} can be recovered, as shown in Fig. \ref{fig:IR_CO2}b.
    
    The lower panel of Fig. \ref{fig:IR_CO2} provides  analogous results when a TRPMD simulation is performed. Compared with the classical results in the upper panel, both the polariton peaks and the \ch{C=O} asymmetric stretch are slightly red-shifted. The broadening of the peaks, as suggested by Ref. \citenum{Lieberherr2023CavMD}, may be attributed to the artifacts in TRPMD rather than a real quantum effect. Since such artifacts have been removed in more recent developments  of TRPMD outside the cavity,\cite{Rossi2018} simulating polaritons using more advanced path-integral approaches is worth exploring. Nevertheless, the path-integral simulation performed here showcases the compatibility of our mesoscale CavMD with path-integral molecular dynamics in the i-PI code, both of which require calling multiple clients to evaluate molecular parameters. Because there is no significant difference between the classical and  TRPMD results, for the calculations below, only classical simulations will be performed.

    \subsubsection{Polariton-polariton scattering beyond the single-mode approximation}

    \begin{figure}
		\centering
		\includegraphics[width=1.0\linewidth]{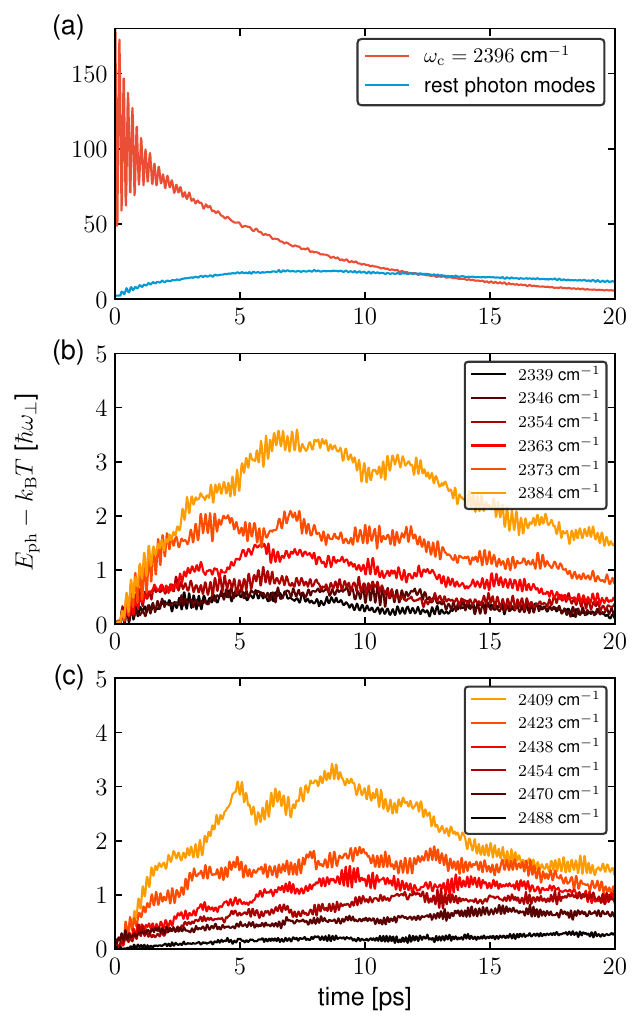}
		\caption{Nonequilibrium photon energy dynamics in the cavity \#1 setup when the cavity mode at $\omega_{\rm c} = 2396$ cm$^{-1}$ ($l_x$ = 12) is initially excited. (a) Photon energy dynamics for the initially excited cavity mode (red line) vs. the summed photon energy of the remaining photon modes with $l_x \neq 12$ (blue line). (b,c) The photon energy dynamics for each individual cavity mode with frequency  (b) $\omega_{\rm c} < 2396$ cm$^{-1}$ or  (c) $\omega_{\rm c} > 2396$ cm$^{-1}$. Here, lines with yellow (black) colors represent cavity modes with frequencies close to (far from) the initially excited photon mode. Energy scattering occurs from the initially excited photon mode to the neighboring cavity modes at different $\omega_{\parallel}$.
		}
		\label{fig:ph_dynamics}
    \end{figure}
    
    Fig. \ref{fig:ph_dynamics}a elaborates further on the 36-mode simulation, when the cavity mode at $l_x = 12$ (corresponding to $\omega_{\rm c} = 2396$ cm$^{-1}$) is strongly excited at $t = 0$. The red line plots the photon energy dynamics of this initially excited cavity mode, while the blue line  plots the summed photon energy of the other cavity modes ($l_x \neq 12$). In general, the excited cavity photon relaxes to the ground state within 10 ps, and the other cavity modes absorb energy transiently up to $t = 5 \sim 10$ ps and then dissipate their energy at later times.  Apart from the energy scattering to the other cavity modes, the relaxation of the $l_x=12$ photon (polaritons) stems also from polariton energy transfer to the dark modes. Cavity loss does not play a role in the photon relaxation dynamics, as the cavity loss is set to zero during our simulations. Please note that since the  photon excitation at $\omega_{\rm c} = 2396$ cm$^{-1}$ is implemented by displacing the photon coordinate at time $t=0$, both the UP and LP have been equally excited. Such an initial excitation can be experimentally achieved by applying a short, wide-band pulse to equally excite both the UP and LP. Hence, the $l_x = 12$ excited photon mode exhibits a strong Rabi oscillation pattern initially ($t < 2$ ps). At later times, however, the Rabi oscillations disappear, as the LP lifetime is shorter than the UP lifetime; \cite{Li2021Relaxation,Li2020Nonlinear} therefore, only the UP signal remains.

    To better understand the dynamics of the remaining photon modes ($l_x \neq 12$), in Figs. \ref{fig:ph_dynamics}b,c, we further provide the photon energy dynamics of each of these photon modes. Lines in yellow represent cavity modes with frequencies close to the initially excited photon frequency ($\omega_{\rm c} = 2396$ cm$^{-1}$), and lines with deeper colors represent cavity modes with frequencies further from it. Clearly, among all of the remaining cavity modes, the closer the frequency to the initially excited photon mode, the more energy the initially unexcited cavity photon receives transiently. This frequency-dependent transient excitation indicates polariton-polariton scattering,  a fundamental mechanism that has been widely studied both experimentally and theoretically in planar Fabry--P\'erot cavities. \cite{Deng2010} 
    Because molecules interact with each other, interactions among polaritons with different $\omega_{\parallel}$ values, which are hybrid light-matter states that can be viewed as molecular excitations modulated by cavity modes, can indeed occur. The polariton-polariton scattering process is unlikely to be mediated by the transiently excited molecular dark modes. This is because the excited dark modes are more likely to transfer energy to those polaritons with frequencies closer to the dark modes, a prediction which is contrary to what is observed in Figs. \ref{fig:ph_dynamics}b,c.
    
    As a short summary, although linear-response polariton spectra seem to suggest that the single-mode approximation is valid, nonequilibrium CavMD simulations emphasize that including $\omega_{\parallel}$-dependent cavity modes is critical for accurately modeling the dynamics of polariton-polariton scattering events.

    \subsection{1D Cavity setups which cannot be used to represent Fabry--P\'erot cavities}
    
    The example of cavity \#1 above represents a set of parameters that could be used to model realistic Fabry--P\'erot cavities. However, it is also useful to investigate under which conditions the numerical scheme cannot be used to simulate realistic Fabry--P\'erot cavities. To this end, we now consider such three additional sets of cavity parameters,  labeled as cavity \#2-\#4.

    \subsubsection{Cavity \#2}
    
    As shown in Table \ref{table:cavity_parameters}, when compared to cavity \#1, cavity \#2 includes  more cavity modes (108 instead of 36) and a smaller $\Delta \omega_x$  (16.67 cm$^{-1}$ instead of 50 cm$^{-1}$). The physical meaning of reducing $\Delta \omega_x$ is demonstrated in Fig. \ref{fig:cavity_setup}: a larger $L_x$ and a  sparser distribution of the molecular grid points along the $x$-direction (the cavity mirror plane). 
    
\begin{figure*}
		\centering
		\includegraphics[width=1.0\linewidth]{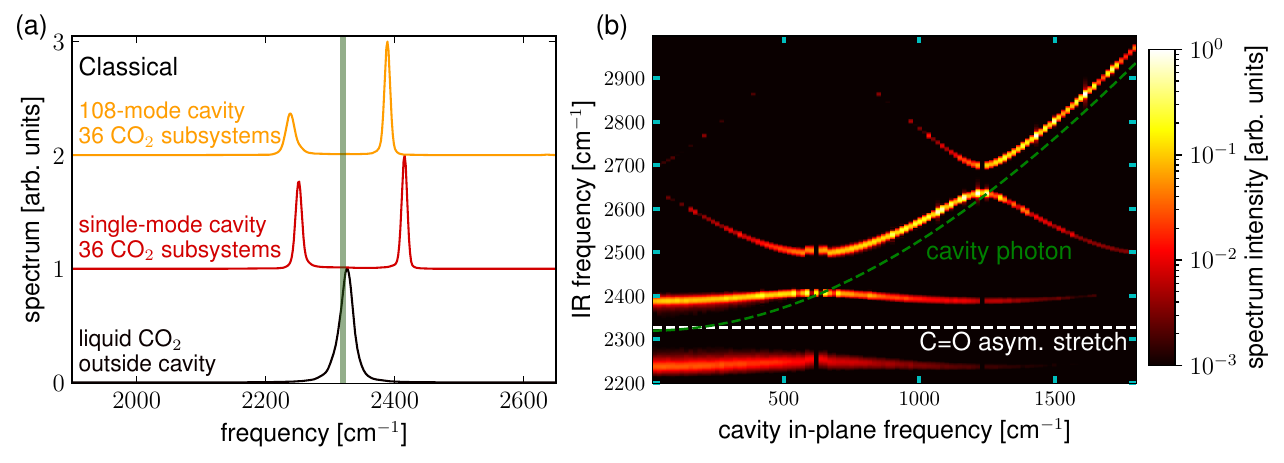}
		\caption{Similar plots as Figs. \ref{fig:IR_CO2}a,b except that  cavity \#1 is replaced by cavity \#2; i.e., 108 (instead of 36) cavity modes are used during the simulation.
		}
		\label{fig:IR_CO2_108modes}
    \end{figure*}

    Fig. \ref{fig:IR_CO2_108modes}a plots the polariton spectrum at normal incidence (yellow line) for this 108-mode cavity.  The polariton spectrum is red-shifted compared to the case when only the lowest-frequency cavity mode (with $l_x = 1$, red line) is included in the simulation. This difference indicates the breakdown of the single-mode approximation. More obviously, as shown in Fig. \ref{fig:IR_CO2_108modes}b, the $\omega_{\parallel}$-dependent polariton spectrum exhibits a much more complicated polariton structure than Fig. \ref{fig:IR_CO2}b: here four polariton branches exist in the  spectrum, with multiple avoided crossings.
    
    \begin{figure*}
		\centering
		\includegraphics[width=1.0\linewidth]{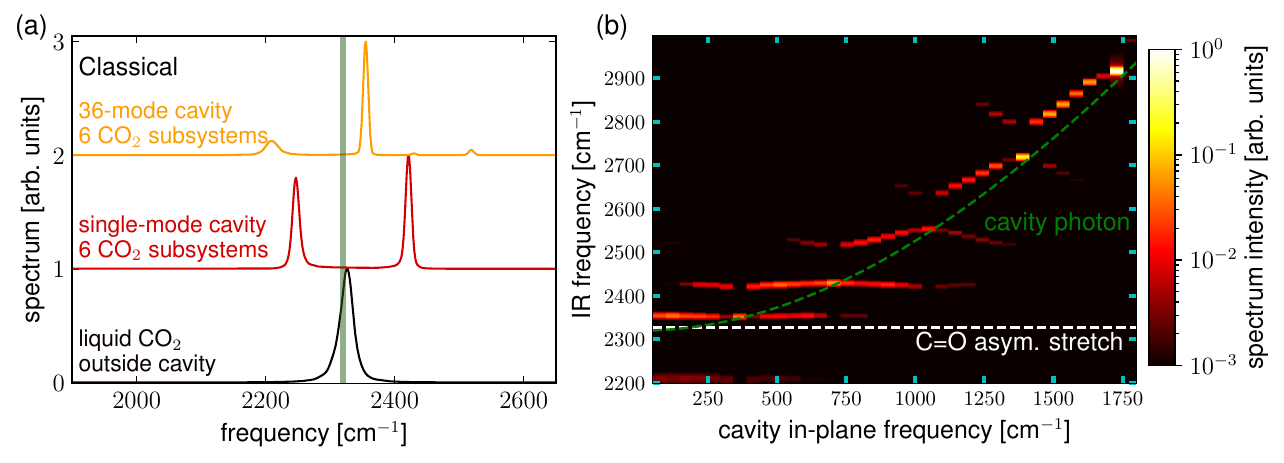}
		\caption{Similar plots as Figs. \ref{fig:IR_CO2}a,b except that here cavity \#1 is replaced by cavity \#3; i.e., only $N_{\rm grid} = 6$ (instead of $N_{\rm grid} = 36$) grid points are used during the simulation and the molecular distribution along the cavity mirror plane (the $x$-direction) becomes very sparse.
		}
		\label{fig:IR_CO2_6b}
    \end{figure*}

     \begin{figure*}
		\centering
		\includegraphics[width=1.0\linewidth]{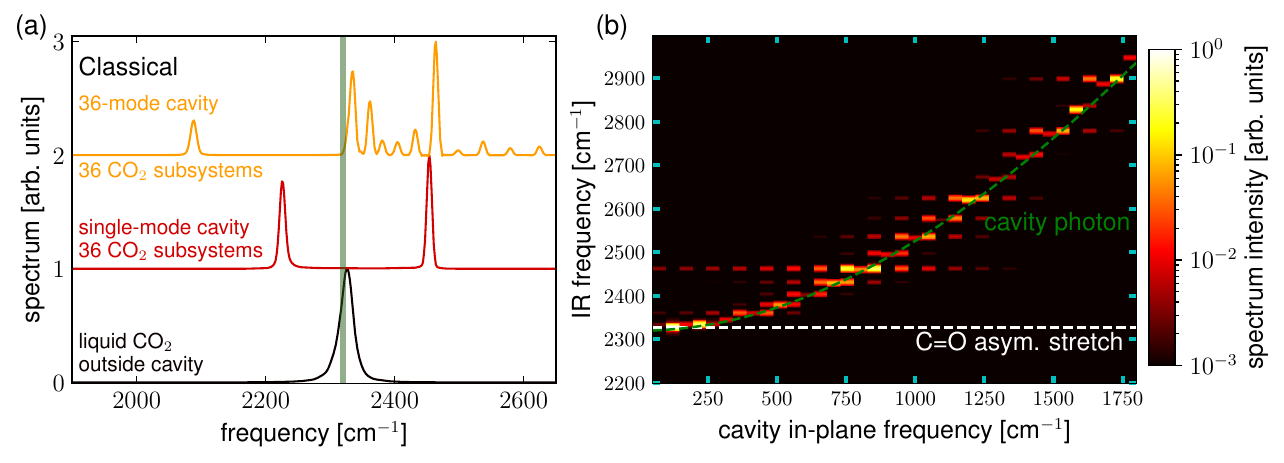}
		\caption{Similar plots as Figs. \ref{fig:IR_CO2}a,b except that here cavity \#1 is replaced by cavity \#4; i.e., the spatial variance of different molecular grid points is eliminated and the long wave approximation is taken.
		}
		\label{fig:IR_CO2_extreme}
    \end{figure*}

    \subsubsection{Cavity \#3}

    For another example, cavity \#3 also resembles cavity \#1 except here a smaller number of molecular grid points, 6 (instead of 36), are evenly distributed. In order to guarantee a similar Rabi splitting as cavity \#1, the effective light-matter coupling $\widetilde{\varepsilon}$ would need to be increased by a factor of $\sqrt{6}$, from $5\times 10^{-5}$ a.u. to $1.225\times 10^{-4}$ a.u.

    Fig. \ref{fig:IR_CO2_6b}a plots the corresponding polariton spectrum at normal incidence (yellow line). This spectrum contains four polariton peaks, which is very different from the case of a single-mode cavity, i.e., when only the lowest-frequency cavity mode ($l_x = 1$) is included in the simulation (red line). Moreover, in the $\omega_{\parallel}$-dependent polariton spectrum (Fig. \ref{fig:IR_CO2_6b}b), cavity \#3 exhibits an even more complicated polariton spectrum than Fig. \ref{fig:IR_CO2_108modes}b, thus strongly indicating significant interaction between cavity photons at different $\omega_{\parallel}$.

    \subsubsection{Cavity \#4}

    Lastly, consider an extreme example --- cavity \#4. Here, all molecular grid points are accumulated at the middle of the cavity mirror plane (along the $x$-direction), while all the other parameters remain the same as cavity \#1. This cavity setup could be regarded as a naive generalization of the single-mode Pauli--Fierz Hamiltonian to the multi-mode case. For this cavity, the corresponding polariton spectrum, as illustrated in Fig. \ref{fig:IR_CO2_extreme}, becomes very complicated. As shown in Fig. \ref{fig:IR_CO2_extreme}a, many peaks coexist in the polariton spectrum at normal incidence (yellow line). Moreover, in the $\omega_{\parallel}$-dependent polariton spectrum (Fig. \ref{fig:IR_CO2_extreme}b), any well-defined polariton branch or avoided crossing completely disappears. Hence, cavity \#4 fails completely  to mimic the polariton dispersion relation in VSC experiments. This  failure emphasizes the fact that the long-wave approximation (i.e., approximating the whole molecular system as localized at a single point) fails when including many $\omega_{\parallel}$-dependent cavity modes.

    \subsection{Cavity \#5: a 2D cavity setup}
    
    After demonstrating the results for several 1D cavities, we found that only the parameter set for cavity \#1 can be used to represent Fabry--P\'erot cavities. However, given that most VSC experiments thus far have employed 2D Fabry--P\'erot cavities,  it is critical to further evaluate the capabilities of our code by directly simulating 2D cavities. After all, the photonic density of states for 1D cavities differs significantly from that for 2D cavities.

    As shown in Fig. \ref{fig:cavity_setup} and Table \ref{table:cavity_parameters}, cavity \#5 extends the parameter set of cavity \#1 to 2D, i.e., with a 2D molecular grid of size $36\times 36$ (1296 grid points). For such a 2D cavity setup,  the cavity frequencies can take
    \begin{equation} \label{eq:omega_c_2d}
        \omega_{\text{c}}(l_x,l_y) = \sqrt{\omega_{\perp}^2 + l_x^2 \Delta \omega_{x}^2 + l_y^2 \Delta \omega_{y}^2} ,
    \end{equation}
    where $\omega_{\perp} = 2320$ cm$^{-1}$, $\Delta \omega_{x,y} = 50$ cm$^{-1}$, and $l_{x,y} = 1, 2, \cdots, 18$.

\begin{figure*}
		\centering
		\includegraphics[width=1.0\linewidth]{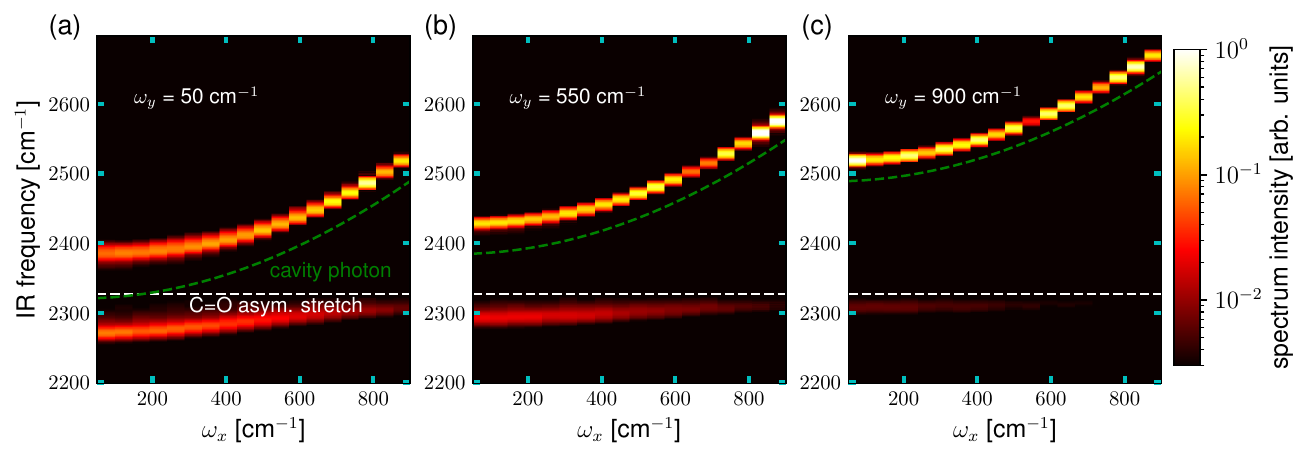}
		\caption{The polariton dispersion relations for cavity \#5 (a 2D cavity) as a function of $\omega_x \equiv l_x\Delta\omega_x$, the in-plane frequency along the $x$-direction, when $\omega_y \equiv l_y\Delta\omega_y$ is chosen as (a) $50$ cm$^{-1}$, (b) $550$ cm$^{-1}$, and (c) $900$ cm$^{-1}$.
		}
		\label{fig:IR_CO2_2D}
    \end{figure*}

    Fig. \ref{fig:IR_CO2_2D} plots the polariton dispersion relation of cavity \#5 after an equilibrium CavMD simulation. We fix $\omega_y \equiv l_y\Delta\omega_y$ and plot $\omega_x \equiv l_x\Delta\omega_x$ for several values of $\omega_y$: (a) $50$ cm$^{-1}$, (b) $550$ cm$^{-1}$, and (c) $900$ cm$^{-1}$. As  $\omega_y$ is increased, the three slices of the polariton dispersion relation show the transition from on-resonance strong coupling at $\omega_x \approx 0$ to  complete off-resonance. Clearly,  cavity \#5 can be regarded as a 2D model  of realistic planar Fabry--P\'erot cavities.
    
    \section{Discussion}\label{sec:discussion}

    After performing simulations on cavity \#1-\#5, we have learned that obtaining a numerical solution to VSC in planar Fabry--P\'erot cavities is nontrivial. On the one hand,
    the computational cost must be bearable in the simulation of Fabry--P\'erot VSC. To this end, both the cavity modes at different $\omega_{\parallel}$ and the molecular distribution need to be discretized. On the other hand, an inappropriate discretization scheme may break the single-mode approximation and produce  inconsistent results when compared with experiments. Hence, one must wonder which discretization scheme is valid.

    \subsection{The role of in-plane translational symmetry in planar Fabry--P\'erot cavities}

    Before discussing how to prepare an appropriate discretization scheme, let us briefly review the reason why the single-mode approximation is valid to describe the linear-response polariton spectrum in planar Fabry--P\'erot cavities. As discussed by Agranovich et al., the polariton eigenvalues in a planar Fabry--P\'erot cavity can be obtained by solving the eigenequation below: \cite{Agranovich2003}
    \begin{equation}\label{eq:polariton_eigenvalues}
        \alpha(\vk_{\parallel}) [ \omega - \omega_{\rm c}(k_{\parallel}) ]
         = \sum_{i} T_i^{\ast}(\vk_{\parallel}) \sum_{\vk_{\parallel}'}T_i(\vk_{\parallel}')\alpha(\vk_{\parallel}')/[\omega - \omega_0] .
    \end{equation}
    Here, $|\alpha(\vk_{\parallel})|^2$ denotes the photonic weight of the polaritons at $\vk_{\parallel}$; $\omega_0$ denotes the molecular excitation frequency; $T_i(\vk_{\parallel})$ denotes the light-matter coupling between the cavity photon mode at $\vk_{\parallel}$ and the $i$-th molecule. Although the explicit form of $T_i(\vk_{\parallel})$ can be multivariate, \cite{Agranovich2003} we are interested only in its dependence on the in-plane phase of the cavity modes. To this end, we can write
    \begin{equation}
        T_i(\vk_{\parallel}) = T_i e^{i\vk_{\parallel} \cdot \vR_{\parallel}^{(i)} } ,
    \end{equation}
    where $ T_i$ denotes the prefactor, and $\vR_{\parallel}^{(i)}$ denotes the molecular position along the cavity mirror plane.  Assuming $T_i$ as a constant and $\alpha(\vk_{\parallel}) \approx \alpha(\vk_{\parallel}')$, we can rewrite Eq. \eqref{eq:polariton_eigenvalues} as
    \begin{equation}\label{eq:polariton_eigenequation_multimode}
        [ \omega - \omega_{\rm c}(k_{\parallel}) ][\omega - \omega_0] \approx \sum_{i}  \sum_{\vk_{\parallel}'} |T_i|^2 e^{-i(\vk_{\parallel} - \vk_{\parallel}')\cdot \vR_{\parallel}^{(i)}} .
    \end{equation}
    Given a homogeneous distribution of molecules along the cavity mirror plane, the summation in Eq. \eqref{eq:polariton_eigenequation_multimode}  over $\vk_{\parallel}'\neq \vk_{\parallel}$ should cancel out with each other. Mathematically speaking, the translational symmetry along the cavity mirror plane gives 
    $\sum_{i}  e^{-i(\vk_{\parallel} - \vk_{\parallel}')\cdot \vR_{\parallel}^{(i)}} \propto \int d \vR_{\parallel}   e^{-i(\vk_{\parallel} - \vk_{\parallel}')\cdot \vR_{\parallel}} \propto \delta(\vk_{\parallel} - \vk_{\parallel}') $, i.e., the summation $\sum_{i}$ can be replaced by a simple integral $\int d \vR_{\parallel}$ only when the molecular distribution is homogeneous along the cavity mirror plane. When $\sum_{i}  e^{-i(\vk_{\parallel} - \vk_{\parallel}')\cdot \vR_{\parallel}^{(i)}}$ is replaced by $\delta(\vk_{\parallel} - \vk_{\parallel}')$,  Eq. \eqref{eq:polariton_eigenequation_multimode} becomes
    \begin{equation}\label{eq:polariton_eigenequation_singlemode}
        [ \omega - \omega_{\rm c}(k_{\parallel}) ][\omega - \omega_0] = \text{const} .
    \end{equation}
    Eq. \eqref{eq:polariton_eigenequation_singlemode} is the polariton eigenequation under the single-mode approximation, i.e., the  cavity modes at different $k_{\parallel}$ effectively do not interact with each other. Note that to derive Eq. \eqref{eq:polariton_eigenequation_singlemode} from Eq. \eqref{eq:polariton_eigenequation_multimode}, we must apply the in-plane translational symmetry.
    
    \subsection{How to satisfy in-plane translational symmetry numerically?}

    As shown above, the (continuous) in-plane translational symmetry is the key to obtain the polariton eigenequation under the single-mode approximation. In our simulations,  both the cavity modes at different $\omega_{\parallel}$ and the molecular distribution are discretized, so the continuous in-plane translational symmetry is of course no longer preserved.

    However, the loss of continuous in-plane translational symmetry  does not preclude numerical simulation of planar Fabry--P\'erot cavities. Instead, we believe that preserving discrete in-plane translational symmetry  is adequate. For our 1D Fabry--P\'erot-like cavity setups, analogous to the cancellation of the $\vk_{\parallel}'\neq \vk_{\parallel}$ terms in Eq. \eqref{eq:polariton_eigenequation_multimode}, it is necessary to reduce the following  error function 
    \begin{equation}\label{eq:check_symmetry}
        \text{Err}(k_x) = \frac{1}{N_{\rm grid}} \sum_{i=1}^{N_{\rm grid}} \sum_{k_x'\neq k_x} \sin(k_x' x_i) \sin(k_x x_i) \rightarrow 0 \text{\ \ \ } 
    \end{equation}
    to zero for any given $k_x$, where $x_i$ denotes the location of the $i$-th grid point.
    
    Given the parameters of the cavity setups in Table \ref{table:cavity_parameters}, in Table \ref{table:cavity_check} we provide the statistics of the error function $\text{Err}(k_x)$ for each 1D cavity setup. While, for cavity \#1,  $\text{Err}(k_x)$ is always exactly equal to zero for arbitrary $k_x$,  for cavities \#2, \#3, and \#4, a finite percentage of  $\text{Err}(k_x)$ has been found to be non-zero for various $k_x$ values. Because we have learned that only cavity \#1 can be used to represent  Fabry--P\'erot cavities employed in experiments, Eq. \eqref{eq:check_symmetry} appears to be a good criterion to check whether the discrete translational symmetry is satisfied for the numerical setups of 1D cavities. Moreover, Table \ref{table:cavity_check} appears to indicate that both the absolute average $\avg{|\text{Err}(k_x)|}$ and the percentage of  $\text{Err}(k_x)\neq 0$ may have a positive correlation with the breakdown of the single-mode approximation: the larger these values, the less the polariton dispersion relation is consistent with experimental findings.

    In summary, to numerically simulate 1D Fabry--P\'erot cavities, which we can use to interpret experiments, we need to guarantee Eq. \eqref{eq:check_symmetry}. Of course, it is also necessary to have a sufficient number of photon modes as well as molecular grid points. Although our discussion above focused on 1D cavities, the conclusion can also be generalized to 2D cavities. It appears that to numerically preserve the translational symmetry of 2D cavities (e.g., cavity \#5), one just needs to make sure the cavity parameters for each dimension preserve the discrete translational symmetry [Eq. \eqref{eq:check_symmetry}] independently.
    
    \begin{table}
    \caption{ Statistics of $\text{Err}(k_x)$ [Eq. \eqref{eq:check_symmetry}] for the 1D cavity setups. Here, the percentage of $\text{Err}(k_x)\neq 0$ is defined as the percentage of  $\text{Err}(k_x)$ satisfying $|\text{Err}(k_x)| >10^{-3}$.
    }
    \label{table:cavity_check}
    \begin{tabular}{lcr}
    \hline
    \hline
    cavity & $\avg{|\text{Err}(k_x)|}$    & \ \ \  percentage of $\text{Err}(k_x)\neq 0$ \\
    \hline
    \#1 & 0.00  & 0\% \\
    \#2 & 0.34  & 35\% \\
    \#3 & 0.42  & 44\% \\
    \#4 & 0.50  & 50\% \\
    \hline
    \hline
    \end{tabular}
    \end{table}

    \subsection{Do cavities in the absence of in-plane translational symmetry have physical meaning?}

    With the above discussion in mind, one may treat cavity \#2-\#4 as a few (unphysical) counterexamples of how  to break the in-plane translational symmetry. However, we also want to emphasize that cavity \#2-\#4 could have real physical meaning. In experiments, if one could prepare cavities with two sets of parallel cavity mirrors along both the $z$- and $x$-directions and then distribute molecules (probably in the solid state) in a similar manner as  in cavity \#2-\#4 (see Fig. \ref{fig:cavity_setup}), one might be able to observe the complicated polariton spectra in Figs. \ref{fig:IR_CO2_108modes}-\ref{fig:IR_CO2_extreme}. These complicated polariton spectra might provide a flexible approach for controlling polariton-polariton scattering and may find wide applications in the future. Even for 2D Fabry--P\'erot cavities used in experiments, by nonuniformly distributing molecules along the cavity mirror plane, one could also observe very complicated polariton dispersion relations. A more comprehensive investigation in this direction is reported elsewhere.\cite{Li2024SymBreakingVSC}

    \subsection{Approaching the macroscopic limit}\label{sec:dis.macro}

    \begin{table*}
    \caption{Parameters for cavity setups of different system sizes. The notation cavity \#1.$X$.$Y$  means that, as compared to cavity \#1 in Table \ref{table:cavity_parameters},  $N_{\rm grid}$ is increased by a factor of $X$ and the cavity mode number ($N_{\rm mode} = l_x^{\rm max}$) is increased by a factor of $Y$. }
    \begin{adjustbox}{width=\linewidth,center}
    \label{table:cavity_parameters_sys_size}
    \begin{tabular}{lccccccccr}
    \hline
    \hline
    cavity   & $\omega_{\perp}$ [cm$^{-1}$] \ & $\Delta\omega_{x}$ [cm$^{-1}$]\ & $\Delta \omega_{y}$ [cm$^{-1}$] \ & $l^{\rm max}_{x}$ \ & $l^{\rm max}_{y}$ \ & $N_{\rm grid}$ \ & $x$ grid positions [$L_x$] \ & $y$ grid positions [$L_y$] \ & $\widetilde{\varepsilon}$ [a.u.]\\
    \hline
    \#1.4.4   & 2320  & 12.5 & 0 & 144 & 1 & 144 & [6.9$\times10^{-3}$, 1.38$\times10^{-2}$, $\cdots$, 0.9931] &  [0.0] & $2.5\times 10^{-5}$ \\
    \#1.16.16   & 2320  & 3.125 & 0 & 576 & 1 & 576 & [1.7$\times10^{-3}$, 3.47$\times10^{-3}$, $\cdots$, 0.9983] &  [0.0] & $1.25\times 10^{-5}$ \\
    \#1.64.64   & 2320  & 0.781 & 0 & 2304 & 1 & 2304 & [4.34$\times10^{-4}$, 8.68$\times10^{-4}$, $\cdots$, 0.9996] &  [0.0] & $6.25\times 10^{-6}$ \\
    \#1.4.1   & 2320  & 50.0 & 0 & 36 & 1 & 144 & [6.9$\times10^{-3}$, 1.38$\times10^{-2}$, $\cdots$, 0.9931] &  [0.0] & $2.5\times 10^{-5}$ \\
    \#1.4.6   & 2320  & 8.33 & 0 & 216 & 1 & 144 & [6.9$\times10^{-3}$, 1.38$\times10^{-2}$, $\cdots$, 0.9931] &  [0.0] & $2.5\times 10^{-5}$ \\
    \#1.4.8   & 2320  & 6.25 & 0 & 288 & 1 & 144 & [6.9$\times10^{-3}$, 1.38$\times10^{-2}$, $\cdots$, 0.9931] &  [0.0] & $2.5\times 10^{-5}$ \\
    \hline
    \hline
    \end{tabular}
    \end{adjustbox}
    \end{table*}
    
    \begin{figure*}
		\centering
		\includegraphics[width=1.0\linewidth]{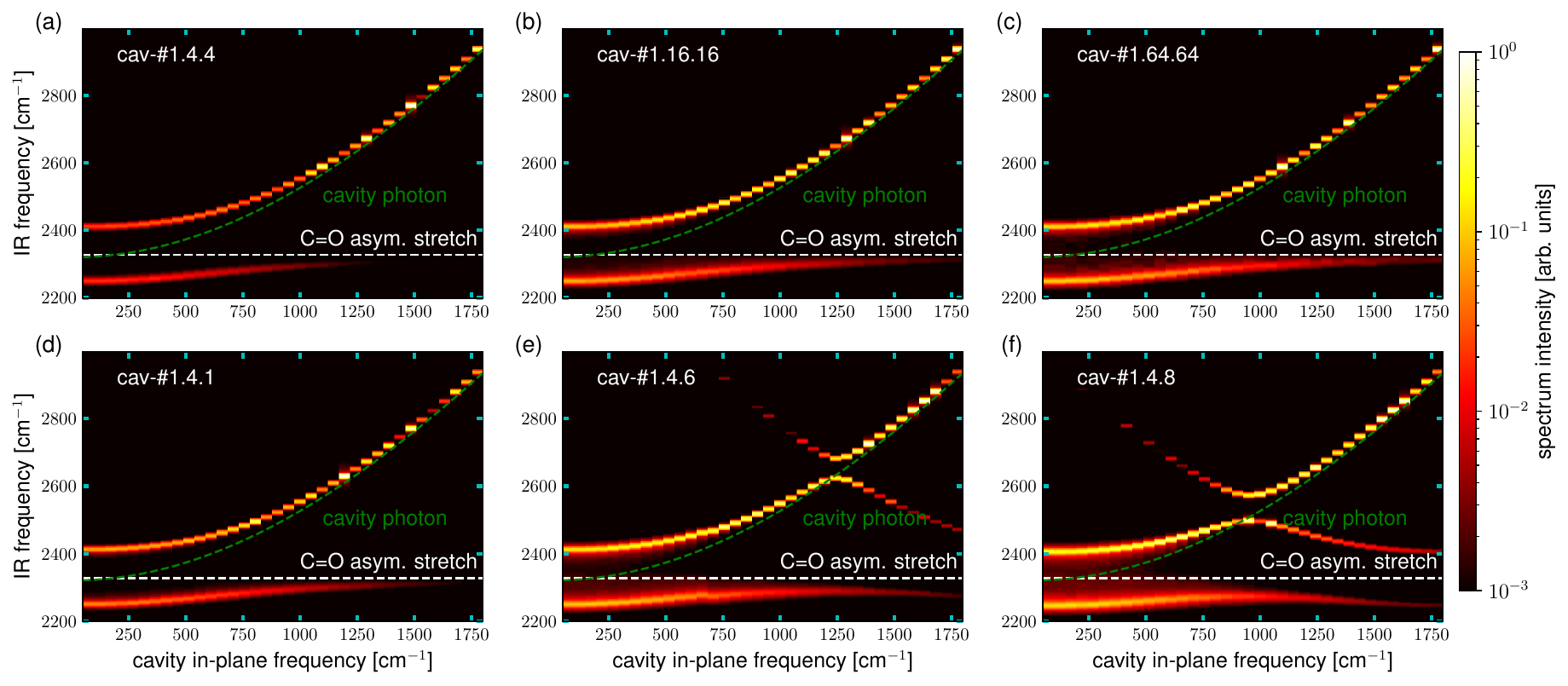}
		\caption{Simulated polariton dispersion relations corresponding to the six cavity setups in Table \ref{table:cavity_parameters_sys_size}. For cavity \#1.64.64 (part c), only 10 equilibrium trajectories were simulated to obtain the spectrum, while for each of the other cavity setups, 40 equilibrium trajectories were simulated.
		}
		\label{fig:IR_CO2_size_dep}
    \end{figure*}

    \begin{figure}
		\centering
		\includegraphics[width=1.0\linewidth]{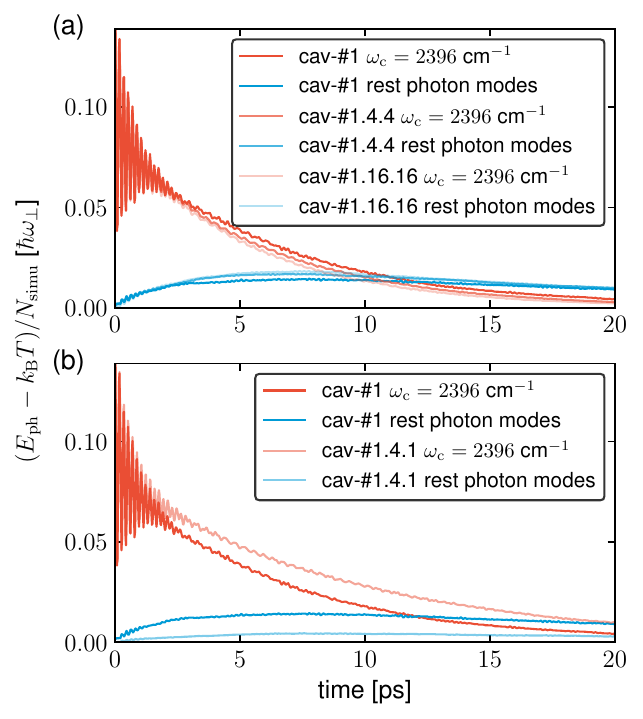}
		\caption{Polariton-polariton scattering dynamics corresponding to a few cavity setups in Table \ref{table:cavity_parameters_sys_size}. The polariton-polariton scattering dynamics for cavity \#1 (plotted in Fig. \ref{fig:ph_dynamics}a) are included in both parts for comparison. For cavities with different system sizes, the cavity mode at the same frequency $\omega_{\rm c} = 2396$ cm$^{-1}$ is initially excited; the initial photon energy is increased proportionally to $N_{\rm grid}$ so that the excitation per molecule [$(E_{\rm ph} - k_B T) / N_{\rm simu}$, the $y$-axis] remains the same.
		}
		\label{fig:ph_dynamics_size_dep}
    \end{figure}

    Finally, we demonstrate how to approach the macroscopic limit in the framework of mesoscale CavMD. To simplify our discussion, we will focus on cavity \#1, which corresponds to a 1D cavity. In order to determine whether the simulation results of cavity \#1  can be used to understand experiments in the macroscopic limit, we will increase the system size and repeat the simulations in Sec. \ref{sec:results}.

    Table \ref{table:cavity_parameters_sys_size} provides six sets of cavity parameters for our simulations, denoted by cavity \#1.$X$.$Y$. Here, for each cavity setup, as compared with cavity \#1, the $N_{\rm grid}$ is increased by a factor of $X$ and the cavity mode number ($N_{\rm mode}$) is increased by a factor of $Y$. When  these two parameters are increased, $\widetilde{\varepsilon} \propto 1/\sqrt{N_{\rm grid}}$ is reduced proportionally to maintain the fixed Rabi splitting. The frequency spacing between the neighboring cavity modes, $\Delta \omega_x = 50.0 \text{\ cm}^{-1}/Y$, is also reduced such that the maximal frequency of the simulated cavity modes remains unchanged.  When $X = Y$, this enlarging process has a physical meaning of increasing the cavity length along the $x$-direction by a factor of $X=Y$ while fixing the molecular density distribution along the cavity mirror plane. When $X \neq Y$, the molecular density distribution along the cavity mirror plane is no longer fixed. All the other simulation details are the same as those in Sec. \ref{sec:simulation_details}.

    Fig. \ref{fig:IR_CO2_size_dep} plots the simulated polariton dispersion relations for the six enlarged cavities listed in Table \ref{table:cavity_parameters_sys_size}. As shown in Figs. \ref{fig:IR_CO2_size_dep}a-c, the polariton dispersion relations remain the same when both $N_{\rm grid}$ and  $N_{\rm mode}$ are increased proportionally ($X=Y=4, 16, 64$). More interestingly, Fig. \ref{fig:IR_CO2_size_dep}d shows that the consistent polariton dispersion relation can also be obtained by increasing only  $N_{\rm grid}$, leaving  $N_{\rm mode}$ unchanged ($X=4, Y=1$). However, Figs. \ref{fig:IR_CO2_size_dep}e,f suggest that setting $Y > X$ leads to polariton dispersion relations breaking the single-mode approximation.

    As discussed previously, numerically preserving the in-plane translational symmetry [or satisfying Eq. \eqref{eq:check_symmetry}] is crucial for validating the single-mode approximation. Because this symmetry should also be preserved as the system size is increased,  Eq. \eqref{eq:check_symmetry} in general limits the possible combinations of $N_{\rm mode}$ and $N_{\rm grid}$. The examples of cavity \#1-\#5 and \#1.$X$.$Y$ appear to suggest that given a fixed $N_{\rm grid}$, ensuring $N_{\rm mode} \lesssim N_{\rm grid}$ can yield reasonable polariton dispersion relations. Although it is unclear whether the upper bound of $N_{\rm mode} \lesssim N_{\rm grid}$ should hold for a general set of 1D parameters, one might  understand this upper bound for our parameters as follows:  When $N_{\rm mode} > N_{\rm grid}$, it becomes impossible to assign every cavity mode to interact with an independent molecular state (as represented by a linearly independent combination of  grid points, or $\sqrt{1/N_{\rm grid}}\sum_i s_i \mathbf{b}_i\sin(k_x x_i)$, where $s_i$ denotes the coefficient and $\mathbf{b}_i$ represents the basis for a single grid point); at least one such a molecular state would have to interact directly with multiple cavity modes, thus violating the single-mode approximation. 

    The polariton-polariton scattering dynamics in Fig. \ref{fig:ph_dynamics_size_dep} provide additional insights when increasing the system size. If $N_{\rm grid}$ and $N_{\rm mode}$ are increased proportionally, as shown in Fig. \ref{fig:ph_dynamics_size_dep}a, the polariton-polariton scattering dynamics remain largely unchanged. 
    However, Fig. \ref{fig:ph_dynamics_size_dep}b demonstrates that polariton-polariton scattering can be greatly suppressed if only $N_{\rm grid}$ is increased and $N_{\rm mode}$ remains the same as in cavity \#1 (cavity \#1.4.1), despite this cavity setup giving reasonable polariton dispersion relations (Fig. \ref{fig:IR_CO2_size_dep}d). Such a suppressed behavior can be understood as follows. The excited polaritons at $\omega_{\rm c} = 2396$ cm$^{-1}$ can transfer energy to two  possible channels: the molecular dark modes with zero photonic contribution, and the polaritons at different in-plane frequencies. When only the $N_{\rm grid}$ is increased, the number of molecular dark modes is increased; because $N_{\rm mode}$ remains unchanged, the number of polaritons in the system remains the same. As a result, when only $N_{\rm grid}$ is increased, the excited polaritons at $\omega_{\rm c} = 2396$ cm$^{-1}$ would be more likely to transfer energy to the dark modes and the polariton-polariton scattering would be greatly suppressed. Overall, one necessary condition to numerically obtain the non-negligible polariton-polariton scattering events in the macroscopic limit is to increase $N_{\rm grid}$ (or more precisely, the number of simulated molecules, $N_{\rm simu}$) and $N_{\rm mode}$ proportionally when enlarging the system size.

    \section{Conclusion}\label{sec:conclusion}

    To summarize, we have provided a  mesoscale CavMD formalism for studying VSC with Fabry--P\'erot-like cavities. The examples of cavities \#1-\#5 showcase the importance of preserving the in-plane translational symmetry for capturing the $\omega_{\parallel}$-dependent polariton dispersion relation observed in experiments. Additionally, our nonequilibrium CavMD simulations have provided initial evidence of elastic vibrational polariton-polariton scattering at nearest neighbor $\omega_{\parallel}$. We have also demonstrated the protocol to approach the macroscopic limit, i.e., by enlarging the system size while reducing the light-matter coupling per molecule (or maintaining the constant Rabi splitting). 

    Our code may serve as an advantageous simulation approach when either the molecular detail (such as the dark modes or chemical reactions) or photonic structure plays crucial roles in the processes of interest. For example, CavMD could be used as a numerical tool to investigate vibrational polariton transport \cite{Suyabatmaz2023} and explore the  experimental conditions necessary to prepare vibrational polariton condensates, \cite{Annir-Sivajothi2022} theoretical hypotheses that have not been experimentally observed. In addition, with a QM/MM description of the molecules,\cite{Li2023QMMM} this code can be used to numerically examine various hypotheses on the origin of the VSC catalytic effects observed in the Ebbesen experiments.

    Moreover, as the formalism of mesoscale CavMD is derived from a general Power--Zienau--Woolley Hamiltonian, our CavMD scheme is gauge invariant and can be easily extended to more complicated photonic environments.\cite{Gautier2022,Voronin2022,Riso2023,Schafer2023,Canales2024,Wright2023,Yoo2021} For instance, by switching the photonic mode functions of a rectangular parallelepiped [Eq. \eqref{eq:mode_function_2d}] to those of a plasmonic nanostructure, plus a QM or QM/MM description of the molecules, one can use this code to study plasmonic catalysis under both weak and strong couplings. Overall, many exciting research opportunities are enabled by mesoscale CavMD. Of course, if one is interested in only linear polariton spectroscopy, it becomes cumbersome to perform expensive molecular dynamics simulations instead of using simplified approaches.\cite{Zhu1990,Li2020Water,Lieberherr2023CavMD}

    In the end, we emphasize that the simulation results for cavities \#2-\#4 also highlight a research opportunity under VSC. Specifically, if the molecular distribution along the cavity mirror plane does not preserve the in-plane translational symmetry, more complicated $k_{\parallel}$-dependent polariton dispersion relations could be experimentally observed. Such a possible avenue is worthy of further investigation, as it provides more flexibility to tune polariton-polariton interactions at different $k_{\parallel}$ values than conventional homogeneous Fabry--P\'erot cavities.


    \section*{Appendix: Including cavity loss}

    \begin{figure*}
		\centering
		\includegraphics[width=1.0\linewidth]{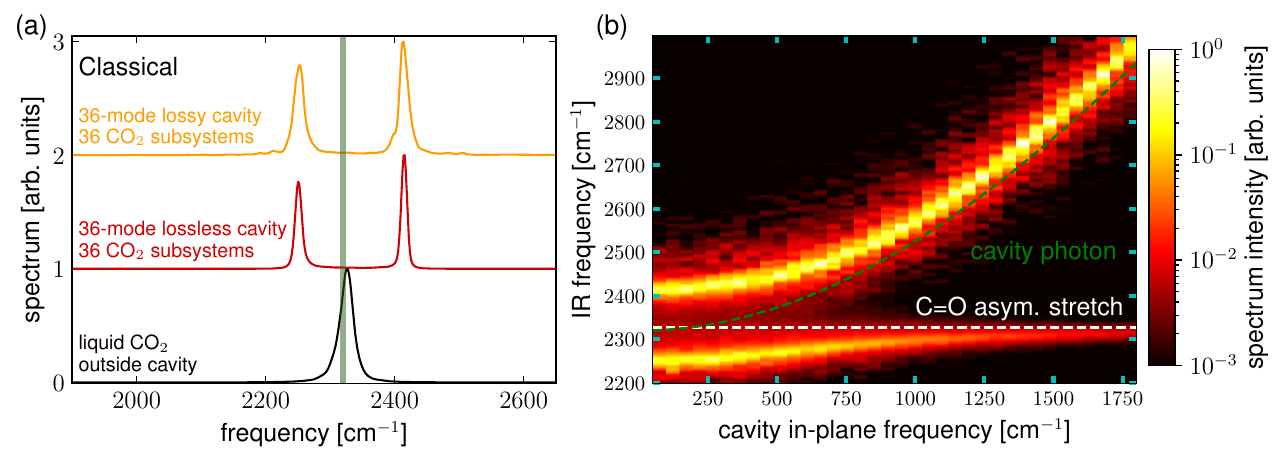}
		\caption{Polariton spectra from classical CavMD simulations. (a) The normalized IR spectrum of liquid \ch{CO2} outside the cavity (black), coupled to a 36-mode lossless cavity (red, Fig. \ref{fig:IR_CO2}a), or a 36-mode lossy cavity with the cavity lifetime as $1/\gamma_{k\lambda}\equiv 300$ fs (yellow).  (b) The polariton spectrum vs the  cavity in-plane frequency $\omega_{\parallel}$ corresponding to the lossy 36-mode cavity. The simulation conditions inside the 36-mode cavity are identical to those in Figs. \ref{fig:IR_CO2}a,b (cavity \#1).
		}
		\label{fig:IR_CO2_lossy}
    \end{figure*}

    The cavity loss has been turned off throughout the calculations  above. As a side note, here we briefly discuss the effects of cavity loss on our simulations. Overall, including cavity loss is very straightforward within the framework of CavMD \cite{Li2021Collective} --- the photonic equation of motion in Eq. \eqref{eq:EOM_new_ph} needed to be replaced by the following form:
    \begin{equation}
        m_{k\lambda} \ddot{\dbtilde{q}}_{k\lambda} = - m_{k\lambda} \omega_{k}^2\dbtilde{q}_{k\lambda} - \widetilde{\varepsilon}_{k\lambda} d_{\text{g},k\lambda}^{\rm sub} - \gamma_{k\lambda} \dbtilde{p}_{k\lambda} + \eta_{k\lambda}(t), 
    \end{equation}
    where $\gamma_{k\lambda}$ and $\eta_{k\lambda}(t)$ denote the damping rate and the Langevin white noise for each cavity mode, respectively. In other words, a standard Langevin thermostat is attached on top of each cavity mode.

    Fig. \ref{fig:IR_CO2_lossy}a plots the polariton spectrum at normal incidence corresponding to the cavity \#1 in Fig. \ref{fig:IR_CO2}a when the cavity lifetime is set as $1/\gamma_{k\lambda}\equiv 300$ fs (yellow line). Here, the polariton linewidth is broadened as compared to the case of the lossless cavity (red line). The polariton dispersion relation in Fig. \ref{fig:IR_CO2_lossy}b also obviously demonstrates the polariton linewidth broadening in the lossy cavity. Note that when the simulation trajectories were used to calculate the polariton spectra in the lossy cavity, only the cavity modes (not molecules) were coupled to the Langevin thermostat.

        \section{ACKNOWLEDGMENTS}
      	This material is based upon the work supported by the start-up funds from the University of Delaware Department of Physics and Astronomy. This research is supported in part through the use of Information Technologies (IT) resources at the University of Delaware, specifically the high-performance computing resources. This work also uses computing resources supported by the AWS Cloud Credit for Research program.

      \section{Date Availability Statement}
      	The code and input files are available on Github at \url{https://github.com/TaoELi/cavity-md-ipi}.
      

    %

\end{document}